\documentclass[a4paper,fleqn,usenatbib]{mnras}

\usepackage{newtxtext,newtxmath}

\usepackage[T1]{fontenc}
\usepackage{ae,aecompl}
\usepackage[utf8]{inputenc}

\usepackage{graphicx}	
\usepackage{amsmath}	

\newcommand{\kms}{$\rm km\,s^{-1}$}
\newcommand{\kmskpc}{$\rm km\,s^{-1}\,kpc^{-1}$}

\title[The kinematic response to the MW spiral structure]
{The kinematic response of RGB centroids and O--B stars to the Milky Way spiral structure: testing the solid-body rotation of the spiral pattern}

\author[P. N. Fedorov et al.] {
P. N. Fedorov,$^{1}$ \thanks{E-mail: pnfedorov@gmail.com (PNF)}
V. S. Akhmetov,$^{1,2,3}$
A. M. Dmytrenko,$^{1,2}$
A. B. Velichko,$^{1}$
D. V. Odynets,$^{1}$
\newauthor
A. P. Zheleznyak,$^{1,4}$
V. P. Khramtsov,$^{1,2}$
S. I. Denyshchenko,$^{1}$
I. B. Vavilova,$^{2}$
D. V. Dobrycheva,$^{2}$ 
\newauthor
   O. Sergijenko,$^{2}$
A. A. Vasylenko, $^{2}$
O. V. Kompaniiets$^{2}$
\\
$^{1}$Institute of astronomy of V.N.Karazin Kharkiv national university, Svobody sq. 4, 61022, Kharkiv, Ukraine \\
$^{2}$Main Astronomical Observatory of the National Academy of Sciences of Ukraine, Akademik Zabolotnyi 27, Kyiv, 03143, Ukraine \\
$^{3}$INAF-Osservatorio Astrofisico di Torino, Via Osservatorio 20, Pino Torinese, Turin, I-10025, Italy \\
$^{4}$Institute of Radio Astronomy of the National Academy of Sciences of Ukraine, 4 Mystetstv Str., Kharkiv, 61002, Ukraine
}

\date{Accepted XXX. Received YYY; in original form ZZZ}

\pubyear{2027}

\begin{document}
\label{firstpage}
\pagerange{\pageref{firstpage}--\pageref{lastpage}}
\maketitle

\begin{abstract}

Using the centroid velocities of RGB stars and the individual velocities of O--B stars from the \textit{Gaia} DR3, we test whether the spiral structure in the investigated region of the Galaxy can be described by a single wave with a constant pattern speed within the framework of Lin--Shu density-wave theory. The amplitudes of the radial and azimuthal velocity perturbations are determined for three centroid samples: centroids within strips 350~pc wide around five spiral arms; centroids within the same strips, with an additional condition on the radial gradient of the radial velocity, $M_{11}$; and all centroids within $5 \leq R \leq 15$~kpc and $130^\circ \leq \theta \leq 270^\circ$. The periodic and smooth non-periodic velocity components are fitted simultaneously. For each sample, the amplitudes derived from the RGB centroid velocities and the individual O--B stellar velocities are used to calculate the relative frequency as a function of Galactocentric radius, $R$, and subsequently the pattern speed at different radii. The analysis is performed for both stellar populations and for two assumed numbers of spiral arms, $m=4$ and $m=5$, using either the amplitude ratio or the radial amplitude alone. None of the cases considered supports a constant pattern speed within the adopted Lin--Shu model. We regard the pattern speed inferred from the radial solution for a specified perturbing potential as our main result. The estimate based on the amplitude ratio provides a complementary result that is consistent with the main conclusion of a non-constant pattern speed, but is less reliable because of the azimuthal response.

\end{abstract}


\section{Introduction}
\label{sec:intro}

The spiral structure of galaxies is of fundamental importance for understanding the dynamics and evolution of disk galaxies. However, the exact nature of spiral arms remains poorly understood. Spiral arms are ubiquitous features of disk galaxies and exhibit considerable morphological diversity across different galactic environments. These structures can be well-defined and continuous over large angular extents (grand-design spirals), fragmented and patchy (flocculent spirals), or exhibit a combination of both characteristics \citep[multi-armed galaxies;][]{Elmegreen1982, Kendall2011}. Despite the important role that spiral arms play in galaxy evolution, explaining their nature remains one of the outstanding problems in astrophysics. To date, numerous theories have been proposed to account for their formation and maintenance, including the density wave theory \citep[][]{Lin1964, Bertin1989, Bertin1996}.

In addition to the density wave theory, other models have also been proposed, such as the swing amplification mechanism \citep[][]{Goldreich1965, Julian1966}; bar-driven spiral responses \citep[][]{Athanassoula2012,Kormendy1979, Tremaine1984,Garma-Oehmichen2021}; tidally driven spiral structures resulting from gravitational interactions with companion galaxies \citep[][]{Salo2000a, Salo2000b, Dobbs2010, Purcell2011, Pettitt2017}; and dynamical models in which spiral structures are self-excited and recurrent \citep[][]{DOnghia2013, Sellwood2014, Sellwood2019, Sellwood2022}.

Nevertheless, several fundamental questions remain unresolved. These include the number of spiral arms, $m$, the lack of consensus on estimates of the spiral pattern speed, $\Omega_{\rm p}$, and, consequently, the precise location of the corotation radius, $R_{\rm cor}$.

Furthermore, it has become evident that the appearance of the spiral structure depends on the stellar population under consideration. \citet{Poggio2021}, and \citet{Drimmel2025} showed that young stars may form fragmented, multi-armed structures, whereas older stellar populations exhibit a more clearly defined two-armed spiral pattern \citep[][]{Churchwell2009, Khanna2025}.
Although advances in observational capabilities have enabled a more detailed characterization of the Milky Way's spiral structure, its true morphology and the dynamical processes governing its evolution remain subjects of active debate. Addressing these questions is important not only for understanding 
the dynamics of the Galactic disc but also for characterising the dynamical properties of Milky Way Near Twins
studied, in particular, in series of our recent works \citep[][]{Vavilova2024, Vavilova2026a, Vavilova2026b, Izviekova2026, Kompaniiets2026}. In this context, the radial behaviour of the spiral pattern speed may provide an additional diagnostic for comparative research of disc galaxies.. In this context, knowledge of the behaviour of the Galactic spiral pattern speed as a function of Galactocentric radius may serve as one of the key criteria for identifying such analogues.

In our study, we adopt the density wave model proposed by \citet{Lin1964} and assume the absence of vertical variations
in the velocity field $\langle V_R \rangle$. Within the framework of the classical quasi-stationary spiral density wave theory
\citep{Lin1969}, a single frequency $\omega$ is assumed to exist, implying a single, constant pattern speed $\Omega_p$.

\begin{equation}
    \Omega_p = \frac{\omega}{m},
\end{equation}

where $m$ is the number of spiral arms corresponding to a given mode. The formulae derived within this theory are local linear expressions describing the kinematic response of the stellar disk. Most estimates of the pattern  speed obtained from young objects within this framework refer to a single effective radius of the sample in the solar neighbourhood. In such cases, the local value of the pattern speed is determined \citep[][]{Dias2019, Bobylev2025, Griv2026}, but its constancy across different Galactocentric radii is not investigated.

However, studies by \citet{Castro-Ginard2021} and \citet{Joshi2023}, based on $Gaia$ EDR3 open clusters, revealed a non-constant spiral wave propagation speed. The derived pattern speed $\Omega_p$ was found to decrease with Galactocentric radius, while the individual propagation speeds of nearby spiral arms closely followed the Galactic rotation curve over the range of 18–56 \kmskpc. These findings are supported by modern numerical simulations, which allow for not only long-lived, rigidly rotating spiral patterns but also transient, radially evolving structures whose pattern speeds may be close to the local angular rotation speed of the Galactic disk \citep[][]{Grand2012a, Grand2012b, Baba2013, Roca-Fabrega2013,DOnghia2013}.

The present study uses $Gaia$ DR3 red giant branch (RGB) stars as kinematic tracers of the spiral arms. This approach allows us to employ a large number of stars while simultaneously covering an extensive region of the Galactic plane.

The aim of this work is not to determine a single value of $\Omega_p$, but rather to address the problem in a more rigorous way: to recover, from the local kinematics, the quantity that is related to the relative frequency of the spiral response in the Lin--Shu linear theory, and only then to examine whether the local results are consistent with a single, constant $\Omega_p$. This order of analysis is methodologically important, since otherwise the assumption of rigid rotation of the spiral pattern may inadvertently become a condition built into the data-processing procedure.

The paper is organized as follows. Section \ref{sec:data} describes the data, centroids, and samples used in the analysis. Section \ref{sec:stellar_response} presents the methodology for determining the amplitudes of the harmonic stellar response and the total phase of the logarithmic spiral. It also describes the decomposition of the velocity field into periodic and non-periodic motions and the estimation of uncertainties. Section \ref{sec:whole_pattern} presents the amplitudes and the pattern speed of the spiral structure as a whole, together with a test of its constancy. Section \ref{sec:ind_arms} is devoted to determining the angular speeds of individual spiral arms and comparing them with the overall pattern speed. Section \ref{sec:check_rad_azim_phases} examines the spiral wave in the interarm regions and analyzes the phase relationships. In Section \ref{sec:pattern_radial}, the pattern speed is determined using only the radial amplitude, while Section \ref{sec:pattern_OB} presents the analysis based on the sample of O--B stars. Section \ref{sec:pattern_speed_m4_m5} compares the results obtained for $m=4$ and $m=5$. Finally, the discussion and conclusions are presented in Section \ref{sec:discussion}.

\section{Observational Data and Centroid Samples}
\label{sec:data}

The mean radial and azimuthal velocity fields of stellar centroids, $\langle V_R\rangle$ and $\langle V_\Theta\rangle$, derived from RGB stars in $Gaia$ DR3 \citep{Vallenari2023}, were used in this study. Each centroid represents the local velocity field averaged within a sphere of radius 1 kpc, with sphere centres spaced at intervals of 0.1 kpc. Since neighbouring spheres overlap, the resulting centroid velocities are not statistically independent measurements. The use of centroids reduces the contribution of random stellar motions and allows large-scale variations in the velocity field to be investigated. Using the criteria described by \citet{Fedorov2026}, we selected 4,798,998 RGB stars from $Gaia$ DR3 and constructed 32,924 centroids. Of these, 24,787 lie within the adopted region, $5<R<15$~kpc and $131^\circ<\theta<269^\circ$. Each sphere contained at least 100 stars.

For each centroid, we calculated the mean coordinates $R$ and $\Theta$, the mean velocities $V_R$ and $V_{\Theta}$, the second velocity moments, and the radial component $M_{11}$ of the symmetric velocity-deformation tensor in the Ogorodnikov--Milne model, together with its standard error \citep{Fedorov2023, Akhmetov2024}.

To identify the spiral arms in the Galactic plane, we used logarithmic spiral loci defined by the parameters reported in the work of \citet{Denyshchenko2024} and shown in Table~\ref{tab:sparm_pars}. We considered the Scutum--Centaurus, Sagittarius--Carina, Perseus, and Outer arms. The Local arm was treated as an additional selection region and was not counted as a fifth arm in the $m=4$ model. This treatment of the Local structure applies to the initial calculation with $m=4$. Section \ref{sec:pattern_speed_m4_m5} additionally considers an $m=5$ model, in which the Local arm is included among the five arms. The spatial selection of objects remains unchanged.

\begin{table}
    \centering
    \caption{Spiral arm parameters from \protect\citet{Denyshchenko2024}.}
    \begin{tabular}{l|cc}
    \hline
    Arm                & $i$             & $a_0$         \\
    \hline
    Scutum-Centaurus   & $-12.04\pm0.34$ & $5.49\pm0.01$ \\
    Sagittarius-Carina & $-12.07\pm0.17$ & $6.88\pm0.01$ \\
    Local              & $-12.34\pm0.32$ & $8.72\pm0.02$ \\
    Perseus            & $-12.07\pm0.18$ & $10.47\pm0.01$\\
    Outer              & $-12.43\pm0.18$ & $12.29\pm0.01$\\
    \hline
    \end{tabular}
    \label{tab:sparm_pars}
\end{table}

Around each spiral locus, we defined a strip with a radial half-width of 350~pc, containing centroids that satisfy $|R-R_{\mathrm{arm}}(\Theta)|\leq0.35$~kpc. The strip width is defined by the difference in Galactocentric radius at a fixed azimuth. To extract the spiral harmonics from the velocity field and determine their parameters, we constructed and analysed the three samples listed in Table~\ref{tab:sample}. The first two contain centroids within the spiral-arm strips and differ only in the additional condition imposed on the gradient $M_{11}$: the ``Strips'' and ``Strips $+M_{11}$'' samples, respectively. The third sample named  ``Full sample'' includes all centroids within the investigated region, including those in the interarm regions. The condition $|M_{11}|/\sigma(M_{11})<2$ selects centroids for which the radial gradient of the mean radial velocity is not statistically distinguishable from zero at the adopted threshold. These centroids are treated as kinematic tracers of the spiral arms. The rationale for this criterion is presented in the work of \citet{Denyshchenko2024}.

\begin{table}
    \centering
    \caption{Centroid samples used in this study.}
    \begin{tabular}{lp{5cm}c}
    \hline
      Sample   & Selection criterion: $5<R<15$ kpc & $N$\\
    \hline
    \hline
      Strips   &  All centroids within the $\pm$350 pc strips 
      around the specified lines tracing the spiral arms. & 7877 \\
    \hline
      Strips  $+M_{11}$ & Centroids within the same strips, but satisfying $\lvert M_{11}\rvert/\sigma_{M_{11}} < 2$ & 3611 \\
    \hline
      Full sample & All centroids, without restrictions on the strips or the $M_{11}$ criterion & 24787 \\
    \hline
    \end{tabular}
    \label{tab:sample}
\end{table}

The two strip samples allow us to assess the sensitivity of the results to the kinematic selection criterion while keeping the spatial region unchanged. The full sample is needed to examine how a single wave manifests itself in the inter-arm regions. The centroid velocities are statistically dependent due to the overlap of the centroid spheres. These dependencies are taken into account when estimating the uncertainties using the spatial block bootstrap.

\section{Determination of the amplitudes of the harmonic stellar response}
\label{sec:stellar_response}

To determine the angular speed of the spiral pattern, the periodic motions induced by it must first be isolated. Lin--Shu density-wave theory \citep{Lin1964} considers a density wave in the galactic disk and an associated small perturbation to the gravitational potential. The maxima of the density perturbation lie along spiral curves and are observed as spiral arms. The perturbations induced by the galactic spiral density wave in the velocity components $V_R$ and $V_{\Theta}$ are described by periodic functions:
\begin{equation}
    V_R = f_R\cos\chi, ~~\delta V_\Theta = f_\Theta\sin\chi
    \label{Eq:VR}
\end{equation}

Here, $f_R$ and $f_{\Theta}$ are the amplitudes of the harmonic stellar response, which, according to the theory, depend only weakly on Galactocentric radius and are assumed to be constant in this work. The quantities $V_R$ and $\delta V_{\Theta}$ are the radial and azimuthal velocity perturbations relative to the unperturbed kinematic baseline defined by the axisymmetric potential. A maximum of one component coincides with a zero crossing of the other. This $90^\circ$ phase difference is referred to as phase quadrature. The phase varies across the disk according to:
\begin{equation}
    \chi(R,\Theta) = m \left[ \cot\,i\,\ln\left(\frac{R}{R_{\rm ref}}\right) - (\Theta-\Theta_{\rm ref}) \right] +\chi_{\rm ref}
    \label{ieq:chi}
\end{equation}
where $m$ is the azimuthal order of the perturbation, i.e. the number of spiral arms in the adopted model, $i$ is the spiral pitch angle, $R_{\mathrm{ref}}$ and $\Theta_{\mathrm{ref}}$ are the reference coordinates, and $\chi_{\mathrm{ref}}$ sets the phase zero point at the time of observation.

The amplitudes of the velocity response are related to the amplitude of the gravitational potential perturbation, $A(R)$, through the following expressions \citep{Lin1969, Yuan1969, Rohlfs1977a}:

\begin{align}  
  & f_R(R) = \left|\frac{k(R)A(R)}{\kappa(R)}\frac{\nu(R)}{1-\nu^2(R)} F^{(1)}_\nu [x(R)] \right|, \nonumber \\
  &  f_\Theta(R) =  \left|\frac{k(R)A(R)}{2\Omega(R)}\frac{1}{1-\nu^2(R)} F^{(2)}_\nu [x(R)] \right|.
    \label{eq:fR_fTheta}
\end{align}

The reduction factors $F_1$ and $F_2$ in these expressions account for the effects of random stellar motions through the parameter $x$ and depend on the dimensionless frequency $\nu$:
\begin{equation}
F_1(\nu,x) = \frac{1-\nu^2}{x}
\left[1-\frac{\pi\nu}{\sin(\pi\nu)}G(\nu,x)\right],
\end{equation}

\begin{equation}
F_2(\nu,x) = (1-\nu^2)\frac{\pi\nu}{\sin(\pi\nu)}
\left(-\frac{\partial G}{\partial x}\right),
\end{equation}
where
\begin{equation}
    G(\nu,x) = \frac{1}{2\pi}\int_{-\pi}^{\pi} \exp[-x(1+\cos s)]\cos(\nu s)ds, \nonumber
\end{equation}
and its derivative
\begin{equation}
    -\frac{\partial G}{\partial x} = \frac{1}{2\pi}\int_{-\pi}^{\pi}(1+\cos s)\exp[-x(1+\cos s)]\cos(\nu s)ds, \nonumber
\end{equation}
while
\begin{align}
  \nu = \frac{m(\Omega_p - \Omega)}{\kappa}, ~~x = \left(\frac{k\sigma_R}{\kappa}\right)^2,
  |k| = \frac{m}{R\,\tan i},~~m = 4. \nonumber
 \label{eq:parameters}
\end{align}

The dimensionless frequency $\nu$ characterises the motion of stars relative to the spiral pattern in units of the epicyclic frequency $\kappa(R)$

\begin{equation}
    \kappa^2(R) = 4\Omega^2(R)\left[1 + \frac{R}{2\Omega(R)}\frac{d\Omega(R)}{dR}\right]
    \label{eq:kappa}
\end{equation}

The parameter $x$ depends on the radial velocity dispersion. In the linear approximation, the amplitudes $f_R$ and $f_{\Theta}$ are slowly varying functions of $R$ compared with the phase $\chi$. In the present calculations, only two radius-independent values are determined for each sample: the radial and azimuthal response amplitudes describing the periodic motions throughout that sample. This is an additional approximation, since the theory itself does not require the amplitudes to remain constant across the entire disk.

\subsection{Separation of periodic and non-periodic motions}

Observed velocities contain both periodic perturbations and non-periodic motions that vary with Galactocentric radius. These contributions therefore need to be separated. For the radial velocity, a natural reference level is provided by $V_{R,0}=0$, because the mean radial motion vanishes in a stationary axisymmetric model. However, the real Galactic disk may also exhibit smooth radial and azimuthal motions that are unrelated to the influence of the density wave. For both velocity components, we adopt the following decomposition:

\begin{align}
    \langle V_c(R,\Theta)\rangle = B_c(R) + C_C\,\cos\chi
    + S_c\,\sin\chi + \varepsilon_c, ~c\in \{R, \Theta\}
    \label{eq:Vc}
\end{align}

Here, $c$ denotes either the radial or azimuthal velocity component. The function $B_c(R)$ describes the smooth, non-periodic variation of the velocity with Galactocentric radius. The sum of the cosine and sine terms represents the harmonic component, corresponding to periodic motion with the adopted spiral phase. The quantity $\varepsilon_c$ represents the residual after subtracting both contributions.

The circular velocity was not subtracted from the observed azimuthal velocity prior to the analysis. Therefore, $B_\Theta$ includes the mean stellar rotation as well as the smooth radial dependence of the asymmetric drift, i.e. the difference between the mean stellar velocity and the circular velocity. Any part of the asymmetric drift not captured by the smooth radial function may remain in the other terms of the decomposition.

The same spiral phase was adopted for all centroids in the analysis:
\begin{equation}
    \chi = m \left[\frac{\ln(R/R_0)}{\tan i} - (\Theta-\pi)\right], ~~R_0=8.275\,{\rm kpc}, i = 12.1^\circ
    \label{eq:chi_R0}
\end{equation}

The value $m=4$ corresponds to a four-armed periodic perturbation. The five observed structures were used solely for the spatial selection of centroids. Their number does not determine the harmonic order: the selection lines and the phase used for the velocity decomposition serve different purposes. The harmonic component of the velocities was determined assuming $m=4$ and a common pitch angle of $12.1^\circ$. The selection lines were not identified with the lines of constant phase of this harmonic. A comparison confirmed that the two do not coincide. To assess the effect of the adopted number of spiral arms, an additional calculation was performed with $m=5$ and the same pitch angle of $12.1^\circ$ (Section \ref{sec:pattern_speed_m4_m5}). For each value of $m$, the spline and harmonic coefficients were redetermined, followed by recalculation of the reduction coefficients, $\nu(R)$, and $\Omega_p(R)$.

The smooth functions $B_c$ were represented by cubic B-splines. The spline and harmonic coefficients were determined simultaneously. If the observed velocities were smoothed independently beforehand, the smoothing could absorb part of the periodic motion and reduce its amplitude. Simultaneous determination of the coefficients avoids precisely this effect. The amplitude and phase of the harmonic component were calculated as
\begin{align}
    f_c = \sqrt{C^2_c + S^2_c},~\alpha = {\rm atan2}(S_c,C_c),~H_c = f_c{\rm cos}(\chi-\alpha_c)
    \label{eq:fc}
\end{align}

\subsection{Uncertainty estimation}

The uncertainties of the amplitudes and phases were estimated using a spatial block bootstrap. In each of the 500 bootstrap realizations, entire 2 × 2 kpc squares were randomly sampled with replacement. The model coefficients were then redetermined using the adopted smoothing level. The radial and azimuthal components were calculated using the same set of squares. The same paired resampling procedure was used when comparing the amplitudes of the overall pattern and individual spiral arms.

The tables report the estimates obtained from the original sample; the values in parentheses are the 16th and 84th percentiles of the bootstrap distribution of the corresponding quantity, unless stated otherwise. The interval between these percentiles contains the central 68\% of the bootstrap results. It indicates how sensitive the result is to changes in the spatial composition of the sample. This estimate is approximate because the centroid spheres overlap and extend across block boundaries.

For the curves, the percentiles were calculated at each radius. To estimate the decline in angular velocity, the value of $\Omega_p$ at the inner radius was subtracted from that at the outer radius in each bootstrap realization. Both angular velocities were calculated using the same estimate of the radial amplitude or amplitude ratio. The 2.5th and 97.5th percentiles of these differences are reported for the main estimates of the decline. When comparing an individual arm with the overall pattern, or solutions obtained for different values of $m$, the percentiles were likewise calculated after subtracting the two angular velocities within each bootstrap realization.

The rotation curve and its derivative, the velocity dispersion, the geometry, and the mean radii of the samples were kept fixed. In the radial equation, the perturbing potential amplitude was also held fixed. The bootstrap intervals therefore do not include uncertainties in these parameters.

Since the amplitudes are calculated as the moduli of pairs of harmonic coefficients, a positive lower bound of their intervals does not in itself indicate a detection of a harmonic component. The estimate from the original data may lie outside the central interval. In particular, for the azimuthal amplitude of the ``entire sample'', the estimate from the original data was sometimes below the 16th percentile of the bootstrap distribution. The symbol $\dagger$ in the tables indicates that the corresponding estimate is not sufficiently reliable. Such estimates are not used for further physical interpretation.



\section{Amplitudes and rotation speed of the spiral pattern as a whole}
\label{sec:whole_pattern}

\subsection{Amplitudes determined for each sample}

In all calculations aimed at determining the angular velocities of the structures, $f_R$ and $f_\Theta$ were assumed to be constant. The first calculation was performed without separating the centroids into individual spiral arms (``Full sample''). All points belonging to a given sample were used jointly. For the strip samples, this approach allowed us to determine the radial and azimuthal amplitudes induced by a single pattern. When determining the amplitudes for the full sample, inter-arm centroids were also included. All centroids were assigned equal weights in the calculation. The resulting amplitudes and their ratios are presented in Table~\ref{tab:ampl_eachsample}.

\begin{table}
    \centering
    \caption{Amplitudes determined for each sample and their ratio. The amplitudes are given in \kms. The 16th and 84th percentiles of the bootstrap distributions are given in square brackets.}
    \begin{tabular}{l|ccc}
    \hline
      Sample            & $f_R$             & $f_\Theta$        & $q = f_R/f_\Theta$ \\
    \hline
      Strips            & 1.3 [0.8, 1.8] & 1.3 [0.8, 2.8] & 1.0 [0.4, 1.7]  \\
      Strips  $+M_{11}$ & 2.0 [1.5, 2.5] & 2.3 [1.3, 4.6] & 0.8 [0.4, 1,7]  \\
      Full sample       & 0.5 [0.3, 1.0] & $\dagger$0.3 [0.3, 1.1] & $\dagger$1.7 [0.4, 2.3]  \\
     \hline 
    \end{tabular}
    \label{tab:ampl_eachsample}
\end{table}

The value $f = 0.30$ cannot be regarded as a reliably measured non-zero amplitude, and the quoted interval reflects only the scatter of the bootstrap amplitude estimates.
As shown in Table~\ref{tab:ampl_eachsample}, the additional  criterion   $M_{11}$ increases both amplitudes, whereas their ratio changes only slightly, from 0.95 to 0.83. Thus, the kinematic selection has a noticeable effect on the amplitudes of the periodic velocity components but only a weak effect on their ratio, which is used to determine the relative frequency $\nu$.


\subsection{Determining the speed of the spiral pattern as a whole}

The method for determining the pattern speed is based on the amplitude ratio. In this case, when considering the ratio of the positive amplitudes, the potential $A(R)$ cancels out, and the expression for the ratio takes the form:
\begin{equation}
    q \equiv \frac{f_R}{f_\Theta} = \frac{2\Omega}{\kappa}\left|\nu\frac{F_1(\nu,x)}{F_2(\nu,x)}\right|
\end{equation}

We used this equation to determine the dimensionless frequency $\nu$ at each specified radius $R$. The equation is nonlinear with respect to $\nu$ and may have multiple roots. Here, a root refers to an individual value of $\nu$ that satisfies the given amplitude ratio at a specified $R$. In calculations based on the derived amplitudes, we adopted the smallest root in the interval $0 < \nu < 1$. The quantities $F_1$ and $F_2$ were calculated using series expansions of the scaled Bessel functions with 120 terms. Thus, for a disk with known properties,
\begin{align}
&\Omega(R) = \frac{V_c(R)}{R},\nonumber\\
&\kappa^2(R) = 2\Omega^2(R) \left[1 + \frac{R}{V_c(R)}\frac{dVc}{dR}\right],
\label{eq:Omega_kappa}
\end{align}
and the reduction factors $F_1$ and $F_2$ evaluated at a given radius, which depend on $x$ and on the unknown frequency $\nu$ itself, the amplitude-ratio method makes it possible to determine $\nu$ as a function of $R$ and subsequently calculate the pattern speed as a function of $R$:
\begin{equation}
    \Omega_p = \Omega(R) + \frac{\nu(R)\kappa(R)}{m}.
    \label{eq:Omega_p}
\end{equation}

The Galactic circular velocity and epicyclic frequency were calculated using our derived circular velocity curve, $V_{\mathrm{circ}}(R)$, for an exponential stellar density distribution, and its derivative \citep[][]{Fedorov2026}. The same radial velocity dispersion profile, $\sigma_{R,\mathrm{RGB}}(R)$, from that study was adopted for all three samples. Since this method of amplitude determination yields only positive values, their ratio constrains only the absolute value of $\nu$, but not its sign. Consequently, each derived root $\nu$ corresponds to two possible values of the pattern speed:
\begin{equation}
    \Omega^-_{p,i} = \Omega_i - \frac{|\nu_i|\kappa_i}{m}, ~~ \Omega^+_{p,i} = \Omega_i + \frac{|\nu_i|\kappa_i}{m}
\end{equation}

In the first expression, $\nu < 0$, and the pattern angular speed is lower than the Galactic circular angular speed. In the second case, $\nu > 0$, and the pattern rotates faster. These two cases cannot be distinguished using positive amplitudes alone. The sign can be determined from the phases of the radial and azimuthal motions, provided that their relationship is reliably established. In the present data, this relationship has not been established (see Section~\ref{sec:check_rad_azim_phases}), therefore, both pattern-speed values are reported. The ``$-$'' and ``$+$'' labels used for the pattern speed below refer to these two cases, respectively.

\subsection{Pattern speed at the mean radius of each sample}

As follows from the equation \ref{eq:Omega_p}, $\Omega_{\mathrm{p}}$ represents a local estimate corresponding to a specific Galactocentric radius $R$. For each of our samples, the pattern speed is determined at a single characteristic radius. We adopt the mean Galactocentric radius of the sample centroids as this characteristic radius. The pattern speed is therefore evaluated at the mean radius
\begin{equation}
    \overline{R} = \frac{1}{N}\sum_{j=1}^{N}R_j
    \label{eq:R}
\end{equation}

The mean is calculated by assigning equal weight to each data point. For each sample, the measured amplitudes are retained, while the circular and epicyclic frequencies, velocity dispersion, and reduction factors are evaluated at $R = \overline{R}$. The results are presented in Table~\ref{tab:pattern_speed}.

\begin{table}
    \centering
    \caption{Pattern speed at the mean radius of each sample. The ``$-$'' and ``$+$'' signs correspond to $\nu < 0$ and $\nu > 0$, respectively. The mean radius is given in kpc, while the pattern speeds
$\Omega^-_p$ and $\Omega^+_p$ are given in \kmskpc.}
    \begin{tabular}{l|ccc}
    \hline
       Sample  & $\overline{R}$, & $\Omega^-_p$, & $\Omega^+_p$,  \\
    \hline
      Strips           & 9.1 &          24.4 [24.0,24.8] &          25.6 [25.3,26.1] \\
      Strips $+M_{11}$ & 9.5 &          23.3 [22.8,23.5] &          24.3 [24.0,24.8] \\
      Full sample      & 9.9 & $\dagger$21.6 [21.2,22.4] & $\dagger$23.9 [23.0,24.2] \\
    \hline
    \end{tabular}
    \label{tab:pattern_speed}
\end{table}

Since the mean radii of the three samples differ, the differences between the pattern-speed values in the rows of the table reflect both the differences in the sample amplitudes and the differences in the corresponding radii. A comparison at identical radii is presented below.

If the amplitudes derived from a given sample describe a single rigidly rotating structure, the pattern speed calculated at different locations (radii) within that structure is expected to be the same. However, the pattern speed determined at a single radius alone is insufficient to establish whether the rotation is indeed rigid. It is therefore necessary to examine whether the same pattern speed is obtained at other radii.

To perform this test, we kept the amplitudes of each sample fixed and successively substituted radii of 5 and 15~kpc, together with the corresponding disk parameters, into the relevant equations. For each sample, this yielded pattern-speed estimates at the respective radii for both negative and positive values of $\nu$.

\begin{table*}
    \centering
    \caption{Pattern angular velocities at the boundaries of the investigated radial range, 5 and 15 kpc.}
    \begin{tabular}{l|rrrr}
    \hline
       Sample         &  \multicolumn{1}{c}{$\Omega^-_p$(5)} & \multicolumn{1}{c}{$\Omega^-_p$(15)} & \multicolumn{1}{c}{$\Omega^+_p$(5)} & \multicolumn{1}{c}{$\Omega^+_p$(15)}\\
    \hline
      Strips           &          45.4 [44.6,45.9] &          13.8 [13.3,14.1] &          47.3 [46.8,48.1] &         15.1 [14.7,15.5] \\
      Strips $+M_{11}$ &          45.5 [44.6,46.0] &          13.8 [13.3,14.1] &          47.2 [46.7,48.1] &         15.0 [14.7,15.5]\\
      Full sample      & $\dagger$44.6 [44.1,45.9] & $\dagger$13.3 [12.0,14.1] & $\dagger$48.1 [46.8,48.6] & $\dagger$15.5 [14.7,15.8]\\
    \hline
    \end{tabular}
    \label{tab:pattern_bound}
\end{table*}

Table \ref{tab:pattern_bound} shows that the pattern angular speeds obtained at different radii are not equal: between 5 and 15~kpc, the speed decreases by approximately 31--33~\kmskpc in all six cases. For the two stripe samples, the pattern-speed values corresponding to $\nu < 0$ differ by no more than 0.126~\kmskpc\, at the same radius. Thus, the additional constraint involving $M_{11}$ does not eliminate the radial dependence of the pattern speed.

The absolute values of the dimensionless frequencies at the corresponding radii are relatively small: $|\nu| = 0.046$--$0.125$ for the ``Strips'' sample and $0.040$--$0.110$ for the ``Strips with the $M_{11}$ constraint'' sample. Consequently, the correction term $\kappa|\nu|/4$ at a given radius is small compared with the value of $\Omega(R)$ at the same location and cannot compensate for its variation with radius. The full sample also yields a decreasing pattern speed, although the amplitudes of the periodic motions are determined less reliably in this case.

\begin{figure}
\centering
\includegraphics[width=\columnwidth]{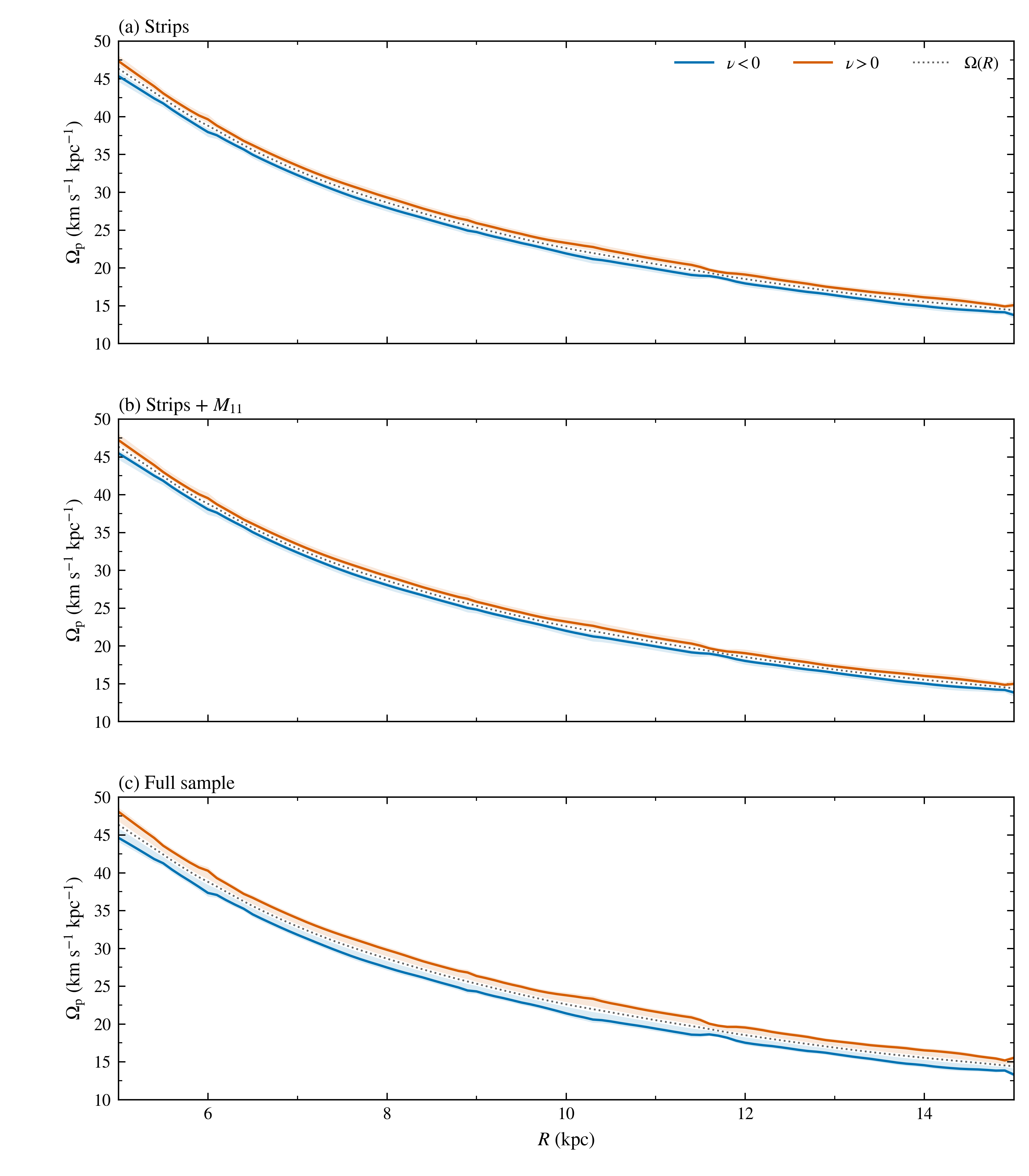}
\caption{Spiral pattern speed as a function of Galactocentric radius for the three samples: (a) Strips, (b) Strips + $M_{11}$, and (c) Full sample. The blue and orange curves correspond to $\nu<0$ and $\nu>0$, respectively. The black dashed curve represents the angular velocity of Galactic circular rotation, $\Omega(R)$. The shaded bands indicate the intervals between the 16th and 84th percentiles of the bootstrap distributions. For each sample, the same amplitude ratio is adopted at all radii.}
\label{fig:pattern_speed}
\end{figure}

\section{Amplitudes and speeds of individual spiral arms}
\label{sec:ind_arms}

\subsection{Determining the amplitudes in each spiral arm}

After analysing the spiral pattern as a whole, the same calculation was performed separately for each of the four main spiral arms. The results for the Local arm are subsequently considered as separate estimates for the region of the Local structure. This allows us to assess the extent to which the motions within individual spiral arms are consistent with the description obtained for the five-arm spiral pattern. The derived amplitudes are presented in Table~\ref{tab:amplitude_arms}. As can be seen from the table, the point estimates of the azimuthal amplitude in some spiral arms are considerably larger than those obtained in the previous analysis. This is not surprising, since the phase varies only slightly within the narrow region of an individual spiral arm for the adopted geometry. Over such a limited range, it is difficult to distinguish a segment of a sinusoidal variation from a smooth radial variation in velocity. This effect is particularly significant for the Outer arm. Therefore, a large amplitude derived for an individual arm does not necessarily imply a stronger wave perturbation. It may instead reflect the ambiguity in separating periodic and non-periodic motions.

\begin{table*}
    \centering
    \caption{Amplitudes of individual spiral arms in the two strip-selected samples.}
    \begin{tabular}{l|cccc}
    \hline
      Arm        &  \multicolumn{2}{c}{Strips} & \multicolumn{2}{c}{Strips + $M_{11}$} \\
                 &  $f_R$  & $f_\Theta$  &  $f_R$      & $f_\Theta$           \\
    \hline
    Scutum-Centaurus       &          1.0 [0.8,2.8] & $\dagger$0.6 [0.8,2.8] & 3.6 [2.8,4.6] & 5.0 [3.5,7.2] \\
    Sagittarius-Carina     &          2.4 [1.7,3.8] &          2.9 [1.9,4.7] & 2.4 [2.0,4.6] & 9.0 [4.0,11.0] \\
    Local                  &          2.3 [1.0,3.2] &          3.3 [1.8,4.4] & 3.9 [2.2,5.2] & 5.1 [3.1,6.6] \\
    Perseus                & $\dagger$0.2 [0.5,1.5] &          1.2 [0.8,2.5] & 2.4 [1.7,3.1] & 3.5 [2.3,5.9] \\
    Outer                  &          1.2 [1.0,2.8] &        10.0 [5.0,13.0] & 3.1 [1.3,5.1] & 13.0 [7.0,16.0]\\
    \hline
    \end{tabular}
    \label{tab:amplitude_arms}
\end{table*}

\begin{figure}
\centering
\includegraphics[width=\columnwidth]{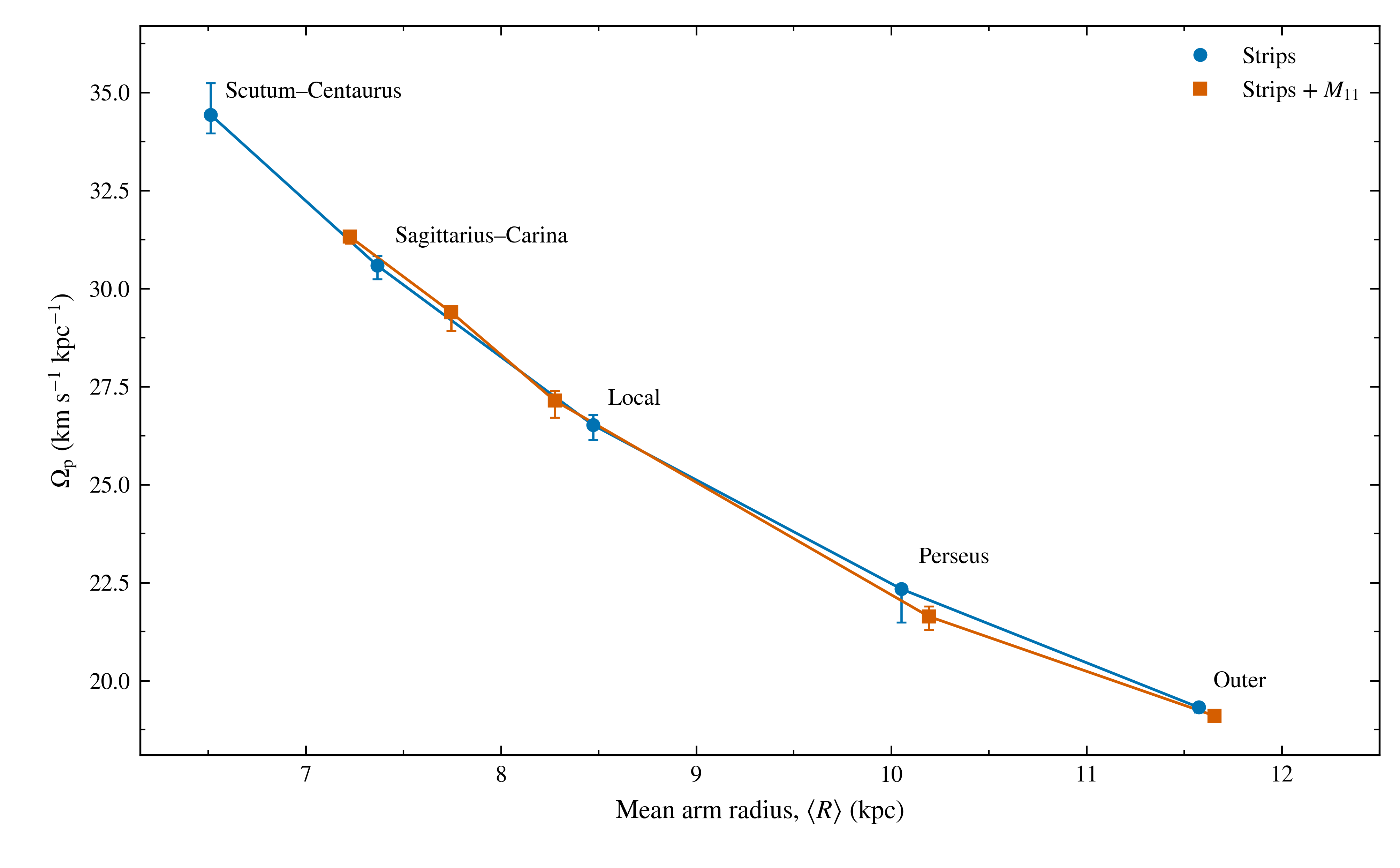}
\caption{Spiral pattern speeds, $\Omega_p$, determined separately for each spiral arm at $\nu<0$ for the Strips and Strips + $M_{11}$ samples. The error bars indicate the intervals between the 16th and 84th percentiles of the bootstrap distributions. The connecting lines are shown for visual guidance only and do not represent fits to the radial dependence. The mean radius of a given spiral arm may differ between the two samples.}
\label{fig:pattern_speed_arms}
\end{figure}

\subsection{Speeds of individual spiral arms and comparison with the result for the overall pattern}

For each spiral arm, $\nu(R)$ is determined from its individual response amplitudes, and the corresponding pattern speed is then calculated at the mean Galactocentric radius of its centroids. For comparison, the pattern speed derived from the amplitudes of the spiral pattern as a whole is evaluated at the same radius. This comparison eliminates differences in radius, as both estimates correspond to the same point on the Galactic rotation curve and the same velocity dispersion. Thus, only the amplitude ratios differ. The results for $\nu<0$ and $\nu>0$ are compared separately.

\begin{table}
    \centering
    \caption{``Strips'' sample. Velocities of individual spiral arms and the pattern at the same mean radius. In each cell, the velocity for $\nu < 0$ is given first, followed by that for $\nu > 0$.}
    \begin{tabular}{l|ccc}
    \hline
      Arm                &  $\overline{R}$ & Arm velocity & Pattern velocity\\
    \hline
    Scutum-Centaurus     & 6.5  &          34.4 [34.0,35.2] & 34.9 [34.4,35.3]  \\
                         &      &          36.6 [35.8,37.1] & 36.2 [35.8,36.6] \\
    Sagittarius-Carina   & 7.4  &          30.6 [30.2,30.8] & 30.5 [30.0,30.9]  \\
                         &      &          31.7 [31.5,32.1] & 31.8 [31.4,32.3]  \\
    Local                & 8.5  &          26.5 [26.1,26.8] & 26.4 [25.9,26.7]  \\
                         &      &          27.4 [27.2,27.8] & 27.6 [27.2,28.1]  \\
    Perseus              & 10.1 & $\dagger$22.3 [21.5,22.3] & 21.8 [21.2,22.2]  \\
                         &      & $\dagger$22.6 [22.7,23.5] & 23.2 [22.8,23.7]  \\
    Outer                & 11.6 &          19.3 [19.2,19.3] & 18.9 [18.6,19.2]  \\
                         &      &          19.4 [19.4,19.6] & 19.8 [19.6,20.2]  \\
    \hline
    \end{tabular}
    \label{tab:strips}
\end{table}

\begin{table}
    \centering
    \caption{``Strips + $M_{11}$'' sample. Velocities of individual spiral arms and the pattern at the same mean radius. In each cell, the velocity for $\nu < 0$ is given first, followed by that for $\nu > 0$.}
    \begin{tabular}{l|ccc}
    \hline
      Arm                &  $\overline{R}$ & Arm velocity & Pattern velocity\\
    \hline
    Scutum-Centaurus     & 7.2  & 31.3 [31.2,31.5] & 31.3 [30.7,31.6] \\
                         &       & 32.3 [32.2,32.5] & 32.4 [32.1,32.9]  \\
    Sagittarius-Carina   & 7.8  & 29.4 [28.9,29.4] & 29.0 [28.4,29.3]  \\
                         &       & 29.8 [29.8,30.3] & 30.2 [29.9,30.8] \\
    Local                & 8.3  & 27.1 [26.7,27.4] & 27.1 [26.5,27.4]  \\
                         &       & 28.2 [27.9,28.6] & 28.2 [27.9,28.8]  \\
    Perseus              & 10.2 & 21.6 [21.3,21.9] & 21.5 [20.9,21.9]  \\
                         &       & 22.7 [22.5,23.1] & 22.8 [22.5,23.5]  \\
    Outer                & 11.7 & 19.1 [19.0,19.2] & 18.9 [18.5,19.1]  \\
                         &       & 19.3 [19.3,19.4] & 19.6 [19.4,19.9]  \\
    \hline
    \end{tabular}
    \label{tab:stripsM11}
\end{table}

As shown in Tables~\ref{tab:strips} and \ref{tab:stripsM11}, the angular pattern speeds derived for the spiral pattern as a whole and for individual spiral arms agree closely at the same Galactocentric radius, although the estimates are based on different amplitude ratios.

Fig.~\ref{fig:pattern_speed_arms} shows that the estimated spiral pattern speeds decrease from the inner to the outer spiral arms in both samples. At the same Galactocentric radius, the estimates obtained for the overall spiral pattern and individual arms agree closely, with differences not exceeding $0.58~\mathrm{km\,s^{-1}\,kpc^{-1}}$ for the ``Strips'' sample and $0.40~\mathrm{km\,s^{-1}\,kpc^{-1}}$ for the ``Strips $+M_{11}$'' sample.

In accordance with the radial variation of the perturbed velocity amplitudes allowed by the Lin–Shu theory, we did not impose equal amplitudes for all spiral arms. Applying the kinematic response equations to the observed motions in individual arms yields estimates of the angular velocity that decrease with Galactocentric radius. Thus, at the mean Galactocentric radius of each arm, when the amplitudes are determined independently, the assumption of a constant angular velocity of a single spiral pattern is not supported.

\section{Consistency Check of the Radial and Azimuthal Motion Phases}
\label{sec:check_rad_azim_phases}

The phase relationship between the radial and azimuthal velocity components is predicted by theory. The radial and azimuthal velocity perturbations are expected to be shifted by a quarter of a period, such that the maximum of one component coincides with a zero crossing of the other. The expected phase shift is $90^\circ$. Unlike the amplitudes, this condition tests the relative spatial arrangement of the perturbed velocity components.

No phase shift was imposed when determining the coefficients of the harmonic function. The phases of the radial and azimuthal velocity perturbations were determined independently, without imposing a quadrature condition. To test for quadrature, the phase difference was reduced to the interval from $0^\circ$ to $180^\circ$, with an expected value of $90^\circ$. Phase differences close to $0^\circ$ and $180^\circ$ are equally far from quadrature. For the ``Strips'' sample, the phase difference was $47.92^\circ$, corresponding to a deviation of $42.08^\circ$ from quadrature. Thus, the derived radial and azimuthal velocity perturbations do not reproduce the expected quarter-period phase shift. Applying the centroid selection criterion $|M_{11}|/\sigma_{M_{11}}<2$ yields a phase difference of $63.82^\circ$, with a deviation of $26.18^\circ$ from quadrature. The $M_{11}$ selection brings the estimated phase difference closer to the expected $90^\circ$, but does not eliminate the discrepancy. For the full sample over $5<R<15$ kpc, the phase difference is $175.52^\circ$, corresponding to a deviation of $85.52^\circ$ from quadrature. Formally, the radial and azimuthal velocity perturbations are therefore nearly in antiphase.

The validity of these conclusions for the ``Strips'' samples is mainly limited by the reliability of the azimuthal component determination. Therefore, the numerical value of the phase discrepancy cannot be regarded as a precisely established property of the wave itself. For the ``Full sample'', it was not possible to identify radial and azimuthal harmonics that simultaneously satisfy the expected phase relationship.

\section{Pattern speed determination using only the radial amplitude}
\label{sec:pattern_radial}

In the previous sections, the radial dependence of the dimensionless frequency $\nu$ was determined from the ratio of the radial and azimuthal velocity amplitudes of the overall spiral pattern, $q=f_R/f_\Theta$.

The advantage of this approach is that the amplitude of the gravitational potential perturbation cancels out. However, the result depends on both measured velocity amplitudes, and the azimuthal amplitude is determined less reliably than the radial amplitude in all calculations. In addition, a significant phase discrepancy between the radial and azimuthal velocity perturbations is also found.

To eliminate the influence of uncertainties in the azimuthal amplitude on the frequency determination, we performed an additional calculation using only $f_R$. A similar approach was adopted, for example, by \citet{Siebert2012}. In the following, we use the same ``Strips'' sample as in the previous sections, consisting of 7877 centroids within strips of radial half-width 350 pc around five spiral structures, without applying the additional $M_{11}$ selection criterion. The radial velocity amplitude of the overall spiral pattern is $1.261~\mathrm{km\,s^{-1}}$, while the individual amplitudes of the spiral arms were obtained from the same joint spline-harmonic analysis and are listed in Table~\ref{tab:amplitude_arms}.

To determine the radial dependence of the absolute value of the dimensionless frequency $\nu$, was determined by solving the following equation:

\begin{equation}
f_R = \left| \frac{kA}{\kappa}\frac{\nu}{1-\nu^2}F_1(\nu,x) \right|, \qquad A = 112.4~(\mathrm{km\,s^{-1}})^2.
\end{equation}

Since the fitted radial response amplitude is non-negative, the radial response equation is solved in terms of its absolute magnitude. Consequently, the amplitude alone does not determine the sign of the relative frequency, and both pattern-speed branches are retained.

Here, $A$ is the adopted amplitude of the gravitational potential perturbation, following the normalization of \citet{Fernandez2008}. The remaining quantities are the same as in the amplitude-ratio calculation: the rotation curve $V_{\mathrm{c,exp}}(R)$ and its derivative, the RGB velocity-dispersion profile, $m=4$, and the adopted pitch angle of $12.1^\circ$. The reduction factor $F_1$ was recalculated for each trial value of $\nu$ at the corresponding $x=(k\sigma_R/\kappa)^2$. The equation was solved numerically over the interval $0<|\nu|<1$. For all radial amplitudes and radii considered, a unique root was found, from which two pattern speeds were calculated: $\Omega_p^-=\Omega-\kappa|\nu|/4$ and $\Omega_p^+=\Omega+\kappa|\nu|/4$.


\subsection{Angular speed of the global spiral structure and individual arms}

We first used the amplitude determined from all 7,877 centroids. At the mean radius of $9.100$~kpc, we obtained $|\nu|=0.7687$, $\Omega_{\mathrm{p}}^{-}=18.474$ and $\Omega_{\mathrm{p}}^{+}=31.563$~km\,s$^{-1}$\,kpc$^{-1}$. We then substituted the amplitudes determined for the individual arms into the equation. The results are presented in Table \ref{tab:rad_control}. For each arm, the disk properties were evaluated at the mean radius of its centroids.

\begin{table*}
    \centering
    \caption{Radial control for the ``Strips'' sample with $A = 112.4;\mathrm{km,s^{-1}}^2$. $R$ is the mean centroid radius in kpc, $f_R$ is in \kms, and velocities are in \kmskpc. The pattern speed for the entire sample was determined using all 7877 centroids, whereas the pattern speed for each spiral arm was determined using only the centroids assigned to that arm.}
    \begin{tabular}{l|ccccc}
    \hline
    Sample             & $R$ & $f_R$ & |$\nu$| & $\Omega^-_p$ & $\Omega^+_p$\\
    \hline
    Whole pattern      & 9.1  &          1.3 [0.8,1.8] &          0.77 [0.63,0.85] &         18.5 [17.8,19.6]  & 31.6 [30.4,32.2] \\
    Scutum-Centaurus   & 6.5  &          1.0 [0.8,2.8] &          0.85 [0.81,0.96] &         25.2 [23.9,25.8] & 45.8 [45.3,47.1]\\
    Sagittarius-Carina & 7.4  &          2.4 [1.7,3.8] &          0.93 [0.90,0.96]&         21.1 [20.8,21.5]  & 41.2 [40.8,41.5] \\
    Local              & 8.5  &          2.3 [1.0,3.2] &          0.91 [0.72,0.94] &         18.6 [18.3,20.3]  & 35.3 [33.6,35.6]\\
    Perseus            & 10.1 & $\dagger$0.2 [0.5,1.5] & $\dagger$0.19 [0.35,0.75] & $\dagger$21.0 [16.5,19.7]  & $\dagger$24.0 [25.3,28.5]\\
    Outer              & 11.6 &          1.2 [1.0,2.8] &          0.64 [0.58,0.86] &          15.6 [14.3,15.9]  & 23.2 [22.8,24.4]\\
    \hline
    \end{tabular}
    \label{tab:rad_control}
\end{table*}

\begin{figure*}
\centering
\includegraphics[width=0.8\textwidth]{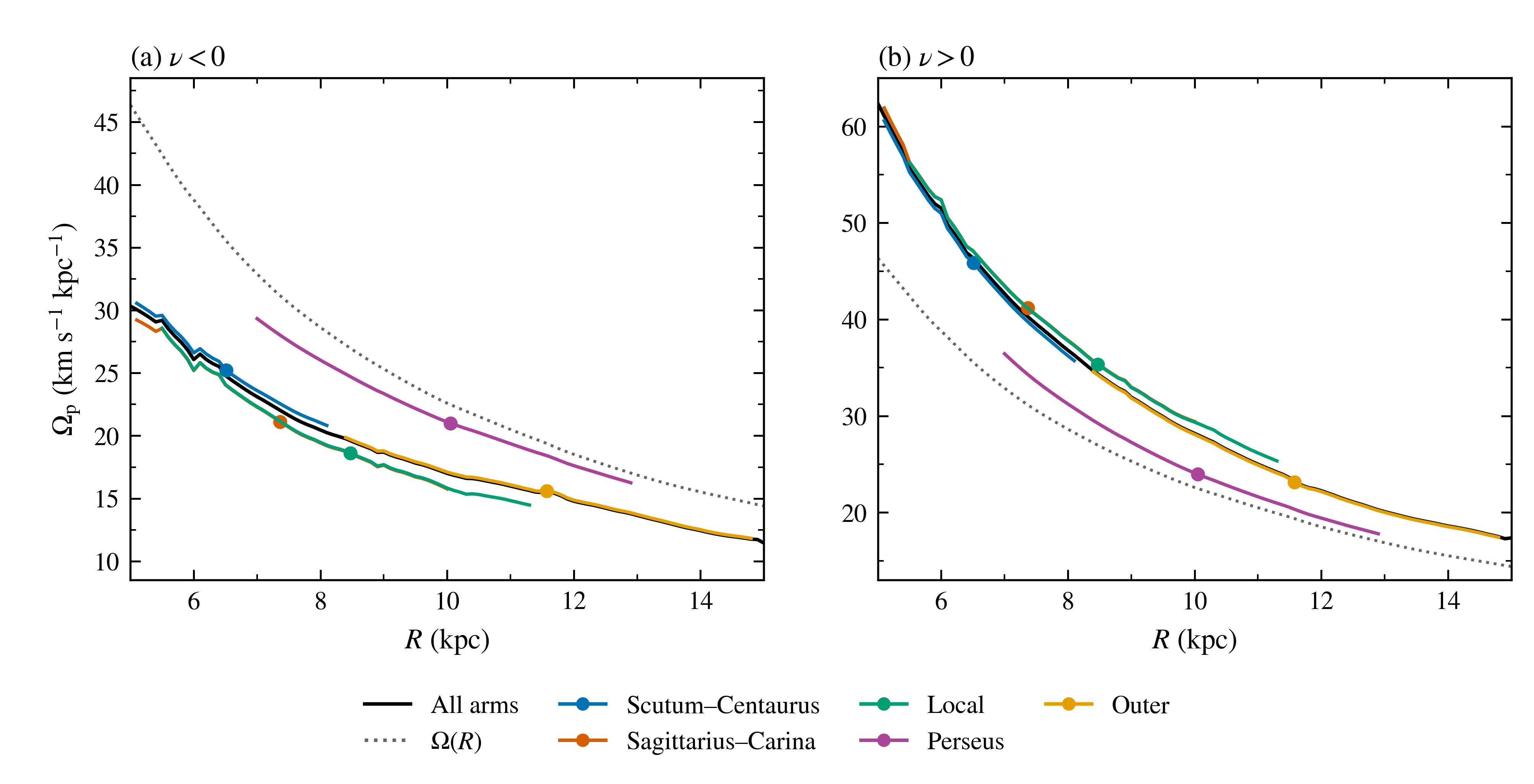}
\caption{Spiral pattern speed, $\Omega_p$, as a function of Galactocentric radius for (a) $\nu<0$ and (b) $\nu>0$, calculated using the radial velocity amplitude of the overall spiral pattern (black curve) and the individual amplitudes of the five spiral arms (coloured curves). Each coloured segment is restricted to the observed radial range of the corresponding arm, with markers indicating the pattern speed at its mean radius. The grey dotted curve represents the angular velocity of Galactic circular rotation, $\Omega(R)$.}
\label{fig:pattern_speed_ind_arms}
\end{figure*}

The angular pattern speed, $\Omega_p$, as a function of Galactocentric radius, calculated using the radial velocity amplitude $f_R$ of the overall spiral pattern and the individual amplitudes of the five spiral arms, is shown in Fig.~\ref{fig:pattern_speed_ind_arms}.

As shown in Fig.~\ref{fig:pattern_speed_ind_arms}, the angular speed of the overall spiral pattern for $\nu<0$ decreases from $30.33~\mathrm{km\,s^{-1}\,kpc^{-1}}$ at $R=5$ kpc to $11.45~\mathrm{km\,s^{-1}\,kpc^{-1}}$ at $R=15$ kpc, while for $\nu>0$ it decreases from $62.37$ to $17.36~\mathrm{km\,s^{-1}\,kpc^{-1}}$. The pattern speeds of the individual spiral arms, with the exception of the Perseus arm, agree closely with those of the overall spiral pattern. The absolute value of the dimensionless frequency, $|\nu|$, lies in the range $0.57$--$0.92$.

\subsection{Comparison with the amplitude-ratio method}

The amplitude-ratio method applied to the ``Strips'' sample yielded $\Omega_p^-=24.406$ and $\Omega_p^+=25.631~\mathrm{km\,s^{-1}\,kpc^{-1}}$ at the mean Galactocentric radius (Table~\ref{tab:pattern_speed}). At the same radius, the radial equation gives values of $18.47$ and $31.56~\mathrm{km\,s^{-1}\,kpc^{-1}}$, respectively. This discrepancy may be related to the adopted normalization of the gravitational potential perturbation. The pattern-speed estimates for individual spiral arms also show considerable variation. However, these differences do not affect the main conclusion of the analysis. Although the use of a fixed potential amplitude $A$ instead of the amplitude ratio leads to substantially different frequencies and pattern speeds, the calculation based solely on $f_R$ does not yield a constant pattern speed over the radial range $5<R<15$ kpc, either for the overall spiral pattern or for individual arms.

Since this approach does not rely on the less reliably determined azimuthal velocity amplitude and does not require determining the phase relationship between the two velocity components, we regard the radial dependence of the dimensionless frequency, $\nu(R)$, derived from the radial velocity amplitude using the adopted normalization of the perturbing gravitational potential following \citet{Fernandez2008}, and the corresponding pattern-speed profile, $\Omega_p(R)$, as the main result of this study. The calculations based on amplitude ratios, as well as comparisons between individual spiral arms and samples, are consistent with this conclusion, although they yield different numerical estimates of the pattern speed.

\section{Pattern angular velocity derived from O--B stars}
\label{sec:pattern_OB}

As shown in the preceding sections, the observational data for RGB stars do not allow us to confirm a constant spiral pattern speed within the framework of the Lin--Shu model. RGB stars have a substantial velocity dispersion, and their kinematics therefore differ appreciably from those of young objects. This may contribute to the inferred variation in the pattern speed.

The aim of this section is to test whether the spiral pattern speed is constant using the observed response of a young stellar population. For this purpose, we use the individual three-dimensional velocities of O--B stars. The spiral-harmonic geometry, rotation curve, and methods for determining the frequency adopted for the RGB samples are retained unchanged for the O--B stars. Thus, we address the same physical question using a different stellar population.

\subsection{Harmonic amplitudes and phase relations}

The initial sample contains 32,358 O--B stars from $Gaia$ DR3. Main-sequence stars were selected using an extinction-corrected colour--absolute magnitude diagram, with $10\,000<T_{\mathrm{eff}}<20\,000$~K and relative parallax uncertainties below $10\%$. The stellar space velocities were calculated in a Galactocentric cylindrical coordinate system using the same solar parameters as for the RGB stars: $R_0=8.275$~kpc \citep{GRAVITY2021} and $(U,V,W)=(11.1,245.8,7.8)$~\kms \citep{Reid2004}. A total of 32,279 stars lie within $5<R<15$~kpc. The spatial region defining the ``Strips'' sample contains 13,804 stars, while the ``Full sample'' contains all 32,279 stars, including 18,475 outside the strips (see Table~\ref{tab:OB_sample}).

\begin{table}
    \centering
    \caption{O–B star samples. $R$ is the Galactocentric radius in kpc, and $R_{\mathrm{mean}}$ is the mean Galactocentric radius of the objects in each sample.}
    \begin{tabular}{l|ccc}
    \hline
     Sample      & $N$    &$R_{\rm cp}$ & Actual range of $R$\\
    \hline
     Strips       & 13,804 & 8.0         & 5.0-12.8  \\
     Full sample & 32,279 & 8.1         & 5.0-12.9 \\
    \hline
    \end{tabular}
    \label{tab:OB_sample}
\end{table}

For both components, we applied the same joint decomposition into a cubic spline $B(R)$ and a harmonic term with the adopted spiral phase as described in Section \ref{sec:stellar_response}.

\begin{table}
    \centering
    \caption{Amplitudes of O--B stars, \kms, and their ratio $q\equiv f_R/f_\Theta$. The 16th--84th percentiles of the bootstrap distributions are given in square brackets. }
    \begin{tabular}{l|ccc}
    \hline
    Sample      & $f_R$ & $f_\Theta$ &  $q$ \\
    \hline
    Strips       & 4.7 	[3.3, 7.4] & 7.9 [4.8, 12.0] & 0.6 [0.3,1.1] \\
    Full sample & 3.1 	[2.5, 4.1] & 1.6 [1.6, 6.0]  & 1.9 [0.5,2.1]  \\
    \hline
    \end{tabular}
    \label{tab:OB_amplitudes}
\end{table}

The jointly estimated stellar response amplitudes in the strips for the radial and azimuthal components are $f_R = 4.69$ and $f_\Theta = 7.86$~\kms, respectively (see Table~\ref{tab:OB_amplitudes}). When the inter-arm regions are included in the analysis, the amplitudes decrease to $f_R = 3.15$ and $f_\Theta = 1.64$~\kms. In this case, all coefficients were redetermined using the full sample. No reliable confirmation of the quadrature relation was obtained for either sample. The large uncertainty in the azimuthal phase and the sensitivity of the azimuthal amplitude to smoothing limit the reliability of frequency estimates based on the amplitude ratio and motivate the use of the independent radial equation.

\subsection{Determination of the pattern speed}

\begin{figure}
\centering
\includegraphics[width=\columnwidth]{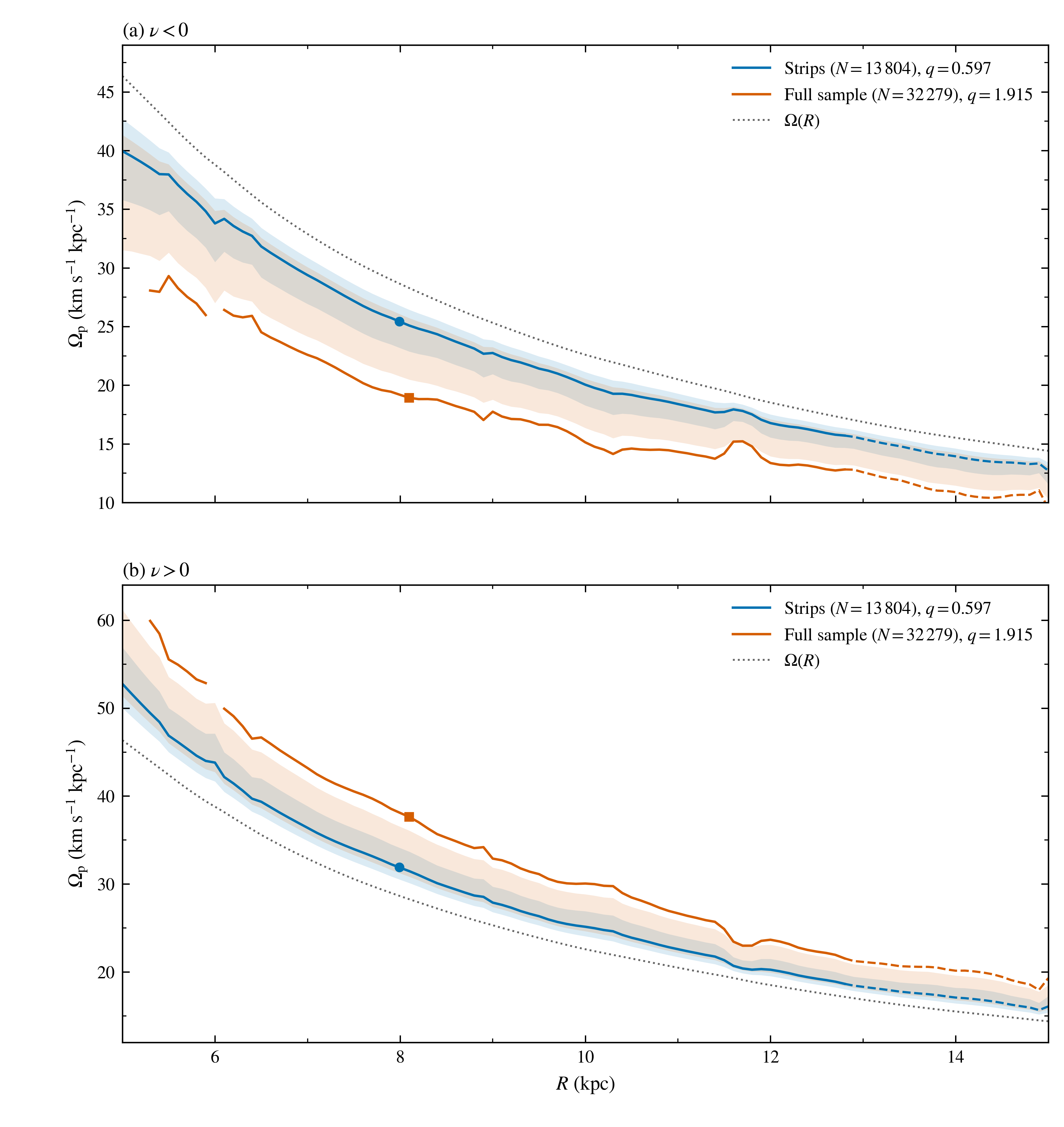}
\caption{Spiral pattern speed, $\Omega_p$, as a function of Galactocentric radius for OB stars, calculated using $\nu(R)$ derived from the amplitude ratio, assuming $\sigma_R=10~\mathrm{km\,s^{-1}}$, $m=4$, and $0<|\nu|<1$. Panels (a) and (b) correspond to $\nu<0$ and $\nu>0$, respectively. The blue and orange curves represent the Strips and Full sample selections, respectively. The markers indicate the mean radii of the samples, while the dashed extensions beyond the outer boundary of the data represent extrapolations. The shaded regions indicate the intervals between the 16th and 84th percentiles of the spatial block bootstrap distributions, calculated using only realizations yielding an admissible root. Gaps in the curves indicate the absence of a solution within $0<|\nu|<1$. The grey dotted curve represents the angular velocity of Galactic circular rotation, $\Omega(R)$.}
\label{fig:ob_pattern_speed}
\end{figure}

We first determined the radial dependence of the frequency $\nu$ using the amplitude ratio obtained for the entire pattern, and then independently using the radial-amplitude equation from Section \ref{sec:pattern_radial} with $A=112.4$~(\kms)$^2$. Both methods were restricted to the interval $0<|\nu|<1$. In calculating the reduction factors for O--B stars, we adopted $\sigma_R=10$~\kms. This is an assumed property of the young population, rather than an observationally determined velocity dispersion. We used the same rotation curve $V_{\mathrm{c,exp}}(R)$, its derivative, $m=4$, and a pitch angle of $12.1^\circ$ as in the preceding calculations. The results evaluated at the mean radii of the samples are presented in Table~\ref{tab:OB_pattern_speed}. Figure~\ref{fig:ob_pattern_speed} shows the spiral pattern speed as a function of Galactocentric radius, calculated using $\nu(R)$ determined from the amplitude ratio.

\begin{table*}
    \centering
    \caption{Pattern speed derived from O--B stars. $R_{\rm cp}$ is in kpc; $\Omega_p$ is in \kmskpc.}
    \begin{tabular}{l|ccccc}
    \hline
     Sample    & Technique & $R_{\rm cp}$ & |$\nu$| & $\Omega^-_p$ & $\Omega^+_p$\\
    \hline
     Strips       & $f_R/f_\Theta$     & 8.0 &          0.32 [0.19,0.55] &          25.4 [23.2,26.8] & 31.9 [30.5,34.1] \\
     Strips       & $f_R, A$ are fixed & 8.0 &          0.60 [0.49,0.72] &          22.7 [21.4,23.7] & 34.7 [33.6,35.9] \\
     Full sample  & $f_R/f_\Theta$     & 8.1 & $\dagger$0.94 [0.26,0.79] & $\dagger$18.9 [20.5,25.7] & $\dagger$37.6 [30.8,36.1] \\
     Full sample  & $f_R, A$ are fixed & 8.1 &          0.48 [0.40,0.56] &          23.5 [22.7,24.3] & 33.0 [32.3,33.9] \\
    \hline
    \end{tabular}
    \label{tab:OB_pattern_speed}
\end{table*}

For both velocity components, we applied the same simultaneous decomposition into a cubic spline $B(R)$ and a harmonic of the adopted spiral phase as in Section \ref{sec:stellar_response}.

The joint fit to the stellar response within the strips yields radial and azimuthal amplitudes of $f_R=4.694$ and $f_{\Theta}=7.858$~km\,s$^{-1}$, respectively. Including the interarm regions reduces these amplitudes to $f_R=3.146$ and $f_{\Theta}=1.643$~km\,s$^{-1}$. In this case, all coefficients were refitted using the full sample. Neither sample provides robust evidence for phase quadrature. The large uncertainty in the azimuthal phase and the sensitivity of the azimuthal amplitude to smoothing limit the reliability of the frequency inferred from the amplitude ratio and motivate the use of the independent radial-amplitude equation.

\begin{figure}
\centering
\includegraphics[width=\columnwidth]{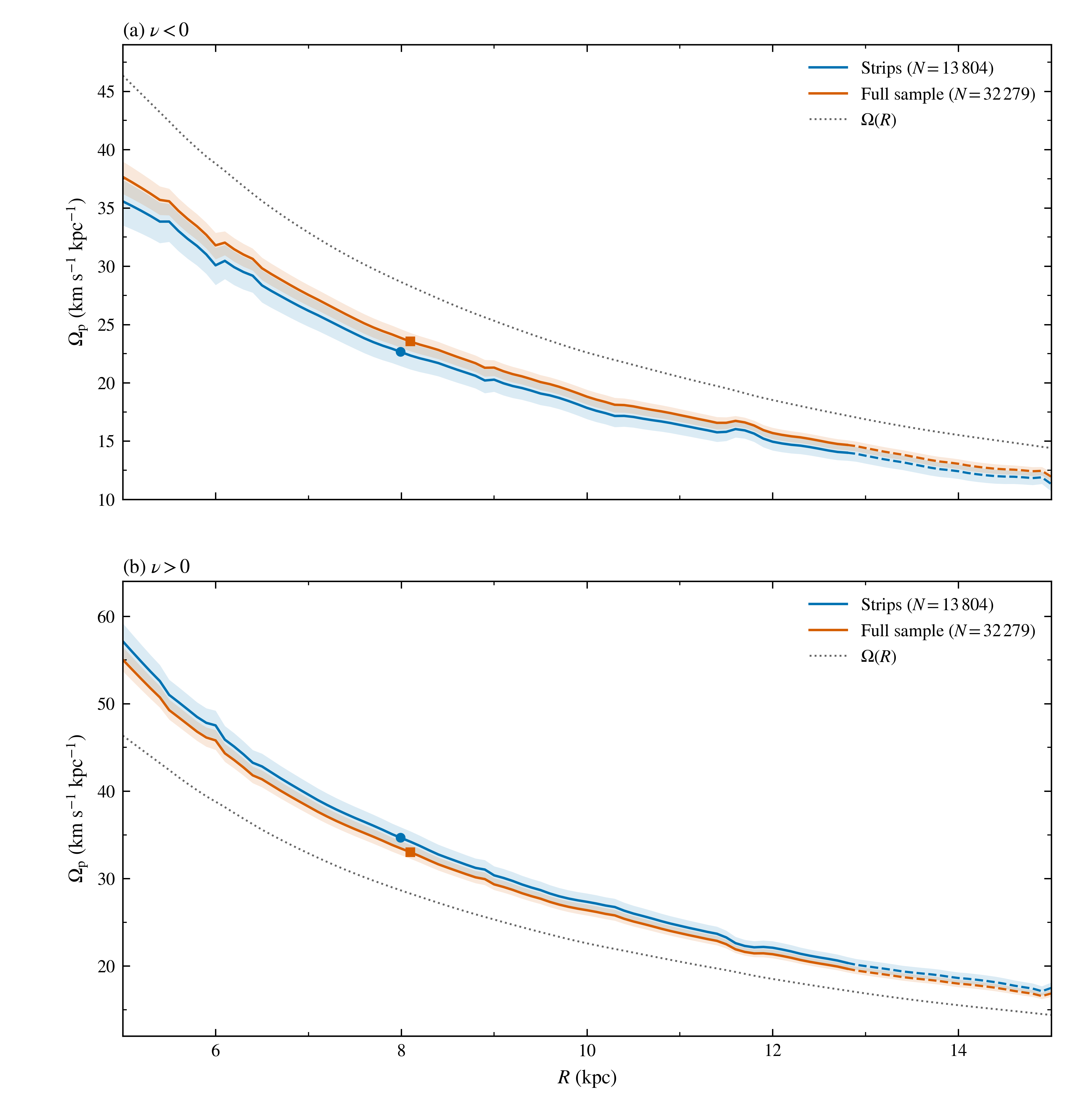}
\caption{Spiral pattern speed, $\Omega_p$, as a function of Galactocentric radius for OB stars, calculated using $\nu(R)$ derived from the radial velocity amplitude. Panels (a) and (b) correspond to $\nu<0$ and $\nu>0$, respectively. The blue and orange curves represent the Strips and Full sample selections, respectively. The shaded regions indicate the intervals between the 16th and 84th percentiles of the bootstrap distributions of $f_R$. The markers indicate the mean radii of the samples, while the dashed extensions beyond the outer boundary of the data represent extrapolations. The grey dotted curve represents the angular velocity of Galactic circular rotation, $\Omega(R)$.}
\label{fig:ob_pattern_speed_radial}
\end{figure}

\subsection{Pattern speed estimates for individual spiral arms}

For the full sample of 32,279 O--B stars, we also performed a separate determination of the amplitudes for the four spiral arms. Each star was assigned to the arm corresponding to the minimum distance in the Galactic plane to one of the five adopted spiral loci. The radial and azimuthal amplitudes were determined independently for each arm. As shown in Table~\ref{tab:OB_vel_amplitudes}, the azimuthal amplitudes are poorly constrained.

\begin{table*}
    \centering
    \caption{Velocity amplitudes in the spiral arm regions. Radii are given in kpc and amplitudes in \kms. The 16th--84th percentiles of the bootstrap distributions are given in square brackets.}
    \begin{tabular}{l|ccccc}
    \hline
      Arm  &$N$ & $\overline{R}$  & $f_R$ & $f_\Theta$ & $q$ \\
    \hline
      Scutum-Centaurus   & 1,116  & 5.9  &            3.9 [3.5, 14.5]   &            32.1 [24.6, 52.0]     & 0.1 [0.1,0.6] \\
      Sagittarius-Carina & 12,014 & 7.2  &            2.7 [2.0, 9.3]    &             8.0 [6.0, 27.4]      & 0.3 [0.2,0.6] \\
      Local              & 16,540 & 8.6  &            3.6 [3.1, 6.0]    &             0.1 [2.6, 13.3]      & $\dagger$24.0 [0.3,1.6] \\
      Perseus            & 2,512  & 10.0 &           14.3 [10.0,20.6]   &             6.9 [5.9, 20.7]      & 2.1 [0.7,2.5] \\
      Outer              & 97     & 11.7 &  $\dagger$72.2 [58.7, 159.3] & $\dagger$160.22 [115.33, 252.33] & $\dagger$0.5 [0.3,1.0] \\
     \hline 
    \end{tabular}
    \label{tab:OB_vel_amplitudes}
\end{table*}

The angular speeds of the spiral arms listed in Table~\ref{tab:ob_arm_pattern_speeds} were calculated using two methods: the amplitude-ratio method and, independently, the radial-amplitude equation.

\begin{table*}
\centering
\caption{Angular pattern speeds of individual spiral-arm regions derived from their own amplitudes using the amplitude-ratio method and the radial-amplitude method with fixed $A$. All values are given in \kmskpc.}
\label{tab:ob_arm_pattern_speeds}
\small
\setlength{\tabcolsep}{4pt}
\begin{tabular}{lcccc}
\hline
Arm & \multicolumn{2}{c}{$f_R/f_\Theta$} & \multicolumn{2}{c}{$f_R$ (fixed $A$)} \\
& $\Omega_p^-$ & $\Omega_p^+$ & $\Omega_p^-$ & $\Omega_p^+$ \\
\hline
Scutum-Centaurus   & 38.3 [34.9,38.5] & 40.2 [39.9,43.6] & 31.5 [27.3,32.0] & 47.0 [46.5,51.1] \\
Sagittarius-Carina & 30.1 [28.7,31.0] & 34.0 [33.0,35.4] & 27.4 [23.6,28.3] & 36.7 [35.8,40.4] \\
Local              & --               & --               & 21.9 [20.5,22.3] & 31.2 [30.8,32.6] \\
Perseus            & --               & --               & 15.8 [15.5,16.3] & 29.2 [28.8,29.6] \\
\hline
\end{tabular}
\end{table*}

Dashes in Table~\ref{tab:ob_arm_pattern_speeds} indicate that no root was found within the adopted frequency interval. For this reason, the main comparison is based on the radial-amplitude equation, which does not involve the azimuthal amplitude.

For the four inner regions, the radial solutions show a systematic decrease in the pattern speed with increasing mean radius for both signs of $\nu$. The outer region contains only 97 stars. The large and uncertain amplitudes obtained for this region do not allow the corresponding pattern speed to be used as a reliable physical estimate.

Thus, for the adopted velocity dispersion and normalization of the perturbing potential, the results obtained from the young stellar population do not support a constant angular speed of the spiral pattern within the framework of the Lin--Shu model used here. We regard the results obtained using the amplitude-ratio method as supplementary. They are consistent with the main conclusion that the pattern speed is not constant, but are less reliable because of the azimuthal response.

\begin{figure*}
\centering
\includegraphics[width=0.8\textwidth]{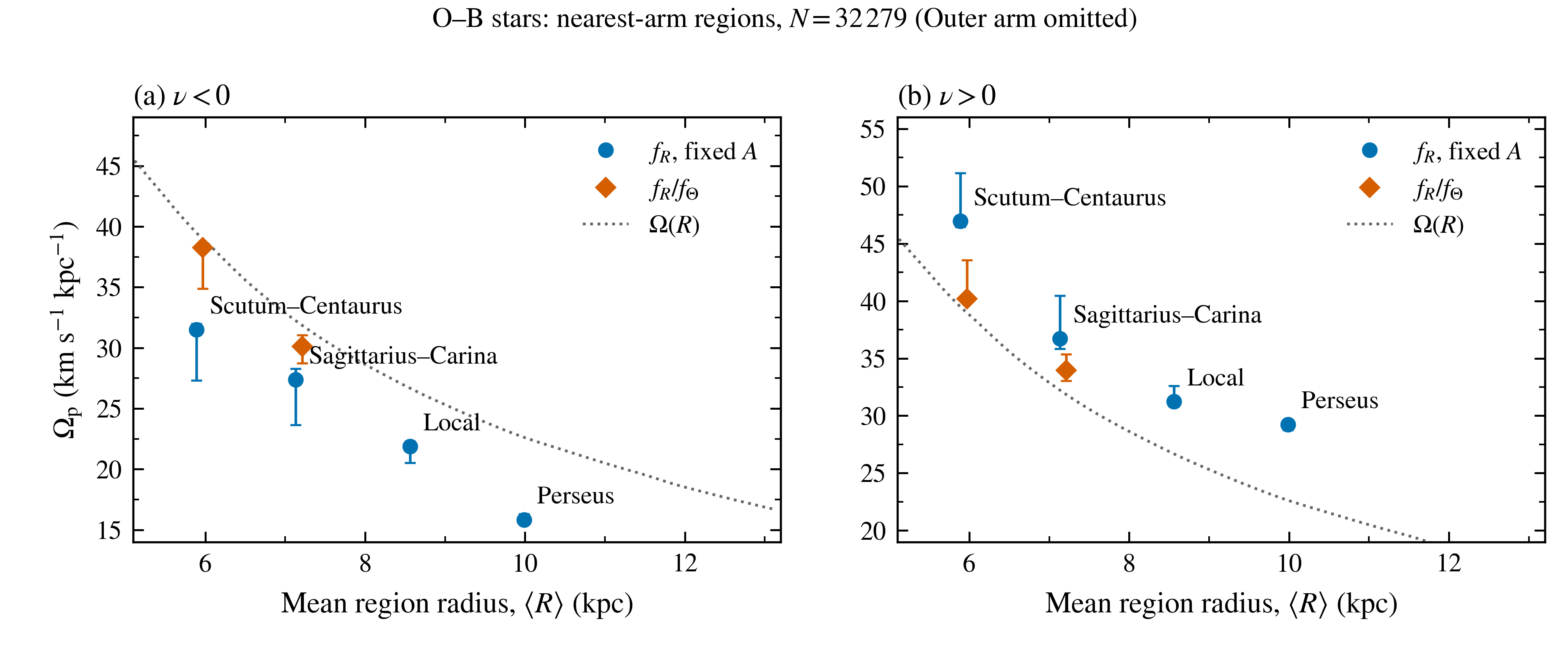}
\caption{Spiral pattern speeds, $\Omega_p$, of individual spiral arms derived from OB stars in the nearest-arm regions ($N=32,279$), excluding the Outer arm. Panels (a) and (b) correspond to $\nu<0$ and $\nu>0$, respectively. Blue circles represent estimates based on the radial velocity amplitude $f_R$ with a fixed $A$, while orange diamonds show estimates obtained from the amplitude ratio $f_R/f_\Theta$. The points are plotted at the mean Galactocentric radii of the corresponding spiral-arm regions. The grey dotted curve represents the angular velocity of Galactic circular rotation, $\Omega(R)$.}
\label{fig:ob_pattern_speed_ind_arms}
\end{figure*}

\subsection{The effect of the radial profile of the perturbing potential on estimates of the pattern angular velocity}

In the previous calculations, the value $A_0=112.4$(\kms)$^2$ of the potential amplitude was adopted based on the normalization by Fernandez. The potential amplitude was assumed to be constant over the entire range of distances considered. To test whether this assumption affects the dependence of the angular velocity on radius, we performed the following test. We adopted the radial form of the potential amplitude proposed by \citet{Eilers2020}. In their model, the exponential variation of the surface density amplitude leads to a dependence of the potential amplitude modulus of the form
\begin{align}
    A(R) \propto R^2\exp(-R/h) \nonumber
\end{align}
For the subsequent calculations, we adopted only the profile shape and the two scale lengths specified by \citet{Eilers2020}, $h = 50$ and $11.36$ kpc. Thus, the comparison of the results obtained with different potential profiles was performed while keeping all other conditions unchanged within each model variant.

\begin{figure*}
    \centering
    \includegraphics[width=0.8\linewidth]{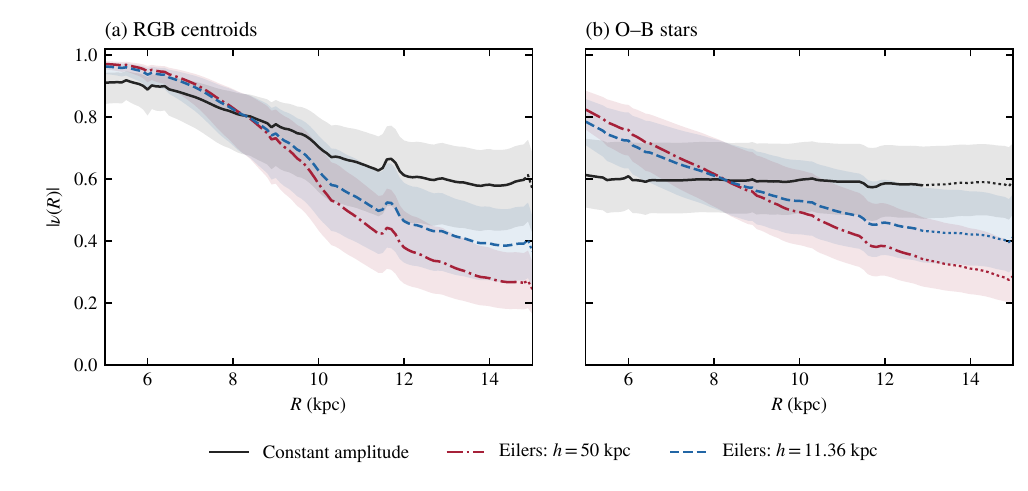}
    \caption{Radial profiles of the modulus of the dimensionless frequency obtained from the radial amplitude of the ``Strips'' sample for $m = 4$.}
    \label{fig:nu-R}
\end{figure*}

Figure~\ref{fig:nu-R} show the radial profiles of the modulus of the dimensionless frequency obtained from the radial amplitude of the ``Strips'' sample for $m = 4$. The upper panels show the RGB centroids, and the lower panels show the O–B stars. The black lines correspond to a constant potential amplitude, while the red and blue lines represent radial profiles with $h = 50$ and $11.36$ kpc, respectively. The shaded regions indicate the 16th–84th percentile ranges of the spatial block bootstrap distributions. All profiles are normalized to the same value of the potential amplitude at the solar radius. 

\begin{figure*}
    \centering
    \includegraphics[width=0.8\linewidth]{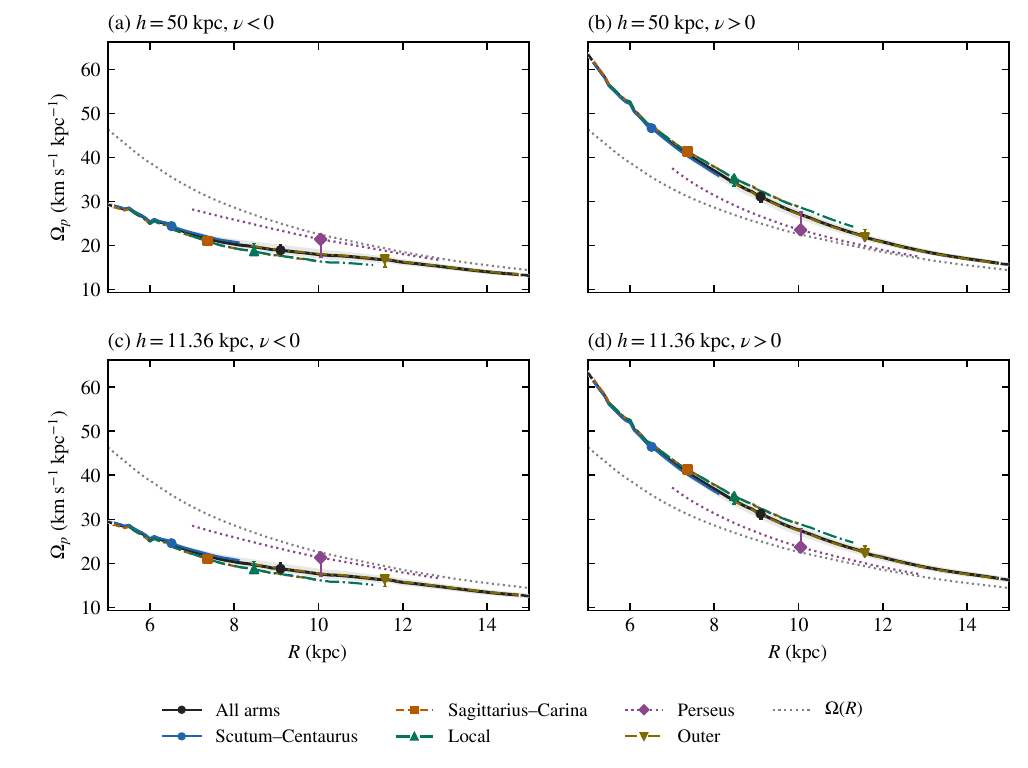}
    \caption{Pattern speed derived from the overall radial amplitude of the pattern (black line) and angular velocities calculated from the amplitudes of individual spiral arms (coloured lines). Each coloured line is shown over the radial range of the corresponding arm. Points indicate estimates at the mean radii; error bars and the gray band represent the 16th–84th percentile bootstrap intervals. The grey dashed line shows the angular velocity of circular motion, $\Omega(R)$. Dashed portions of the black curve indicate extrapolation beyond the sample boundaries.}
    \label{fig:pattern_speed_radial}
\end{figure*}

\begin{figure*}
    \centering
    \includegraphics[width=0.8\linewidth]{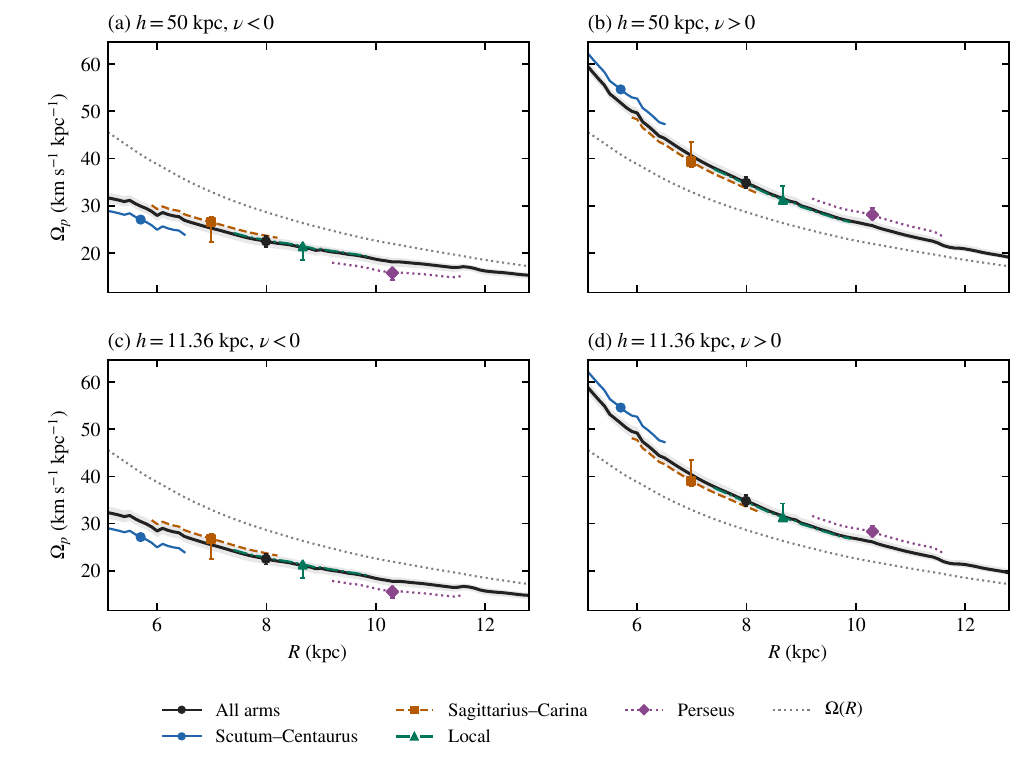}
    \caption{Pattern speed derived from the overall radial amplitude of the pattern (black line) and from the amplitudes of individual spiral arms (coloured lines). Each coloured line is shown over the radial range of the corresponding arm. Points indicate estimates at the mean radii; error bars and the grey band represent the 16th–84th percentile bootstrap intervals. The grey dashed line shows the angular velocity of circular motion, $\Omega(R)$. The outer arm (24 stars) is not shown.}
    \label{fig:pattern_speed-R_OB}
\end{figure*}

As shown in Fig. \ref{fig:pattern_speed_radial}, the pattern angular velocity decreases between 5 and 15 kpc for both signs of $\nu$. For $h = 50$ kpc, it decreases from 29.3 to 13.1 \kmskpc\, and from 63.4 to 15.7 \kmskpc\,; for $h = 11.36$ kpc, it decreases from 29.4 to 12.5 \kmskpc\, and from 63.3 to 16.3 \kmskpc.

The dependence of the pattern speed on Galactocentric radius for the O--B stars is shown in Fig.~\ref{fig:pattern_speed-R_OB}.

Between 5.1 and 12.8 kpc, the pattern angular velocity decreases for both signs of $\nu$. For $h = 50$ kpc, it decreases from 31.7 to 15.3 \kmskpc\, and from 59.5 to 19.1 \kmskpc; for $h = 11.36$ kpc, it decreases from 32.3 to 14.8 \kmskpc\, and from 58.8 to 19.6 \kmskpc.

Thus, for both \citet{Eilers2020} profile scale lengths, the derived angular velocity of a single spiral pattern decreases with radius for both the RGB centroids and the O--B stars. For the RGB centroids, the estimates for individual arms do not form a monotonic sequence: for $\nu < 0$, the estimate for the Perseus Arm is higher than that for the Local Arm.

\section{Testing the dependence of the results on the number of spiral arms}
\label{sec:pattern_speed_m4_m5}
The initial calculation adopts a four-armed harmonic, although five observed structures are used to select the objects. To assess whether this difference affects the main conclusion, we repeated the calculation with $m=5$, including the Local arm. We retained the pitch angle of $12.1^\circ$, the spiral loci and widths of the selection strips, the rotation curve, and the observed velocity-dispersion profile for RGB stars. For O--B stars, we adopted $\sigma_{v_R}=10$~\kms\, in both cases. The change in $m$ was accounted for at every stage of the calculation. As in the main analysis, uncertainties were estimated from 500 spatial block bootstrap resamples using $2\times2$~kpc blocks. The primary comparison was based on the radial-amplitude equation with $A=112.4$~(\kms)$^2$.

\begin{table*}
\centering
\caption{Joint estimates for $m=4$ and $m=5$. In the amplitude, $q$, and $|\nu|$ columns, the values for $m=4$ and $m=5$ are given in that order. The amplitudes are expressed in \kms, the mean Galactocentric radii in kpc, and the angular pattern speeds in \kmskpc. The dimensionless frequencies and pattern speeds were derived from the radial equation.}
\label{tab:pattern_speed_m4_m5}
\setlength{\tabcolsep}{4pt}
\begin{tabular}{lccccccc}
\hline
Sample & $\overline{R}$ & $f_R$ & $f_\Theta$ & $q$ & $|\nu|$ & $m=4$ & $m=5$ \\
& (kpc) & $4/5$ & $4/5$ & $4/5$ & $4/5$ & $\Omega_p^-/\Omega_p^+$ & $\Omega_p^-/\Omega_p^+$ \\
\hline
RGB ``Strips''           & 9.1 & 1.3 [0.8,1.8] & 1.3 [0.8,2.8] & 1.0 [0.4,1.7] & 0.8 [0.6,0.9]& 18.5 [17.8,19.6] & 19.4 [19.0,19.8] \\
                         &     & 1.2 [1.0,1.7] & 2.2 [1.5,3.3] & 0.6 [0.4,0.9] & 0.8 [0.8,0.9] & 31.6 [30.4,32.2] & 30.7 [30.2,31.1] \\
RGB ``Strips $+M_{11}$'' & 9.5 & 2.0 [1.5,2.5] & 2.3 [1.3,4.6] & 0.8 [0.4,1.7] & 0.9 [0.8,0.9] & 16.9 [16.6,17.4] & 18.3 [18.0,18.8] \\
                         &     & 1.5 [1.1,2.1] & 3.1 [1.9,4.6] & 0.5 [0.3,0.9] & 0.9 [0.8,0.9] & 30.6 [30.1,30.9] & 29.3 [28.7,29.6] \\
RGB ``Full sample''      & 9.9 & 0.5 [0.3,1.0] & 0.3 [0.3,1.1] & 1.7 [0.4,2.3] & 0.4 [0.3,0.6] & 19.7 [17.8,20.5] & 20.3 [19.0,21.2] \\
                         &     & 0.4 [0.3,0.7] & 0.3 [0.3,0.9] & 1.4 [0.4,2.2] & 0.4 [0.3,0.6] & 25.8 [24.9,27.7] & 25.2 [24.3,26.5] \\
OB ``Strips''            & 8.0 & 4.7 [3.3,7.4] & 7.9 [5.0,12.0] & 0.6 [0.3,1.1] & 0.6 [0.5,0.7] & 22.7 [21.4,23.7] & 23.7 [23.1,24.5] \\
                         &     & 5.5 [3.9,7.4] & 6.6 [4.0,9.7]  & 0.8 [0.5,1.4] & 0.6 [0.5,0.7] & 34.7 [33.6,35.9] & 33.7 [32.8,34.3] \\
OB ``Full sample''       & 8.1 & 3.1 [2.5,4.1] & 1.6 [1.6,6.0] & 1.9 [0.5,2.1] & 0.5 [0.4,0.6] & 23.5 [22.7,24.3] & 24.9 [24.5,25.5] \\
                         &     & 3.0 [2.3,3.5] & 1.6 [1.2,4.2] & 1.8 [0.7,2.4] & 0.4 [0.4,0.5] & 33.0 [32.3,33.9] & 31.7 [31.1,32.1] \\
\hline
\end{tabular}
\end{table*}

\begin{figure*}
    \centering
    \includegraphics[width=0.8\linewidth]{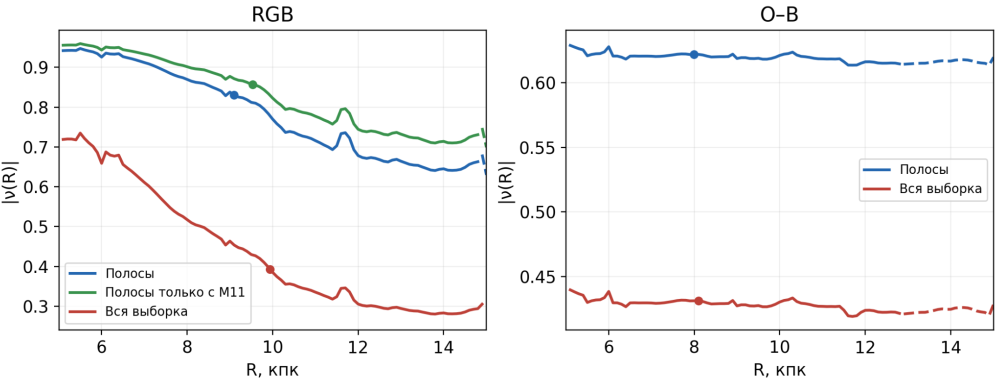}
    \caption{The $|\nu(R)|$ profiles derived from the radial amplitudes for $m = 5$. Left: RGB stars; right: O--B stars. The points mark the mean radii of the samples. The dashed extensions of the O--B curves beyond the outer boundary of the data indicate extrapolation.}
    \label{fig:nu_R}
\end{figure*}

As can be seen from Table~\ref{tab:pattern_speed_m4_m5}, both the amplitudes and the response frequency change for $m = 5$. A separate $\nu(R)$ relation was derived for each amplitude obtained. Although the absolute value of the frequency in the radial solution varies relatively little for young stars, as shown in Fig. \ref{fig:nu_R}, this does not ensure a constant $\Omega_{\mathrm{p}}$.

\begin{figure*}
\centering
\includegraphics[width=0.8\textwidth]{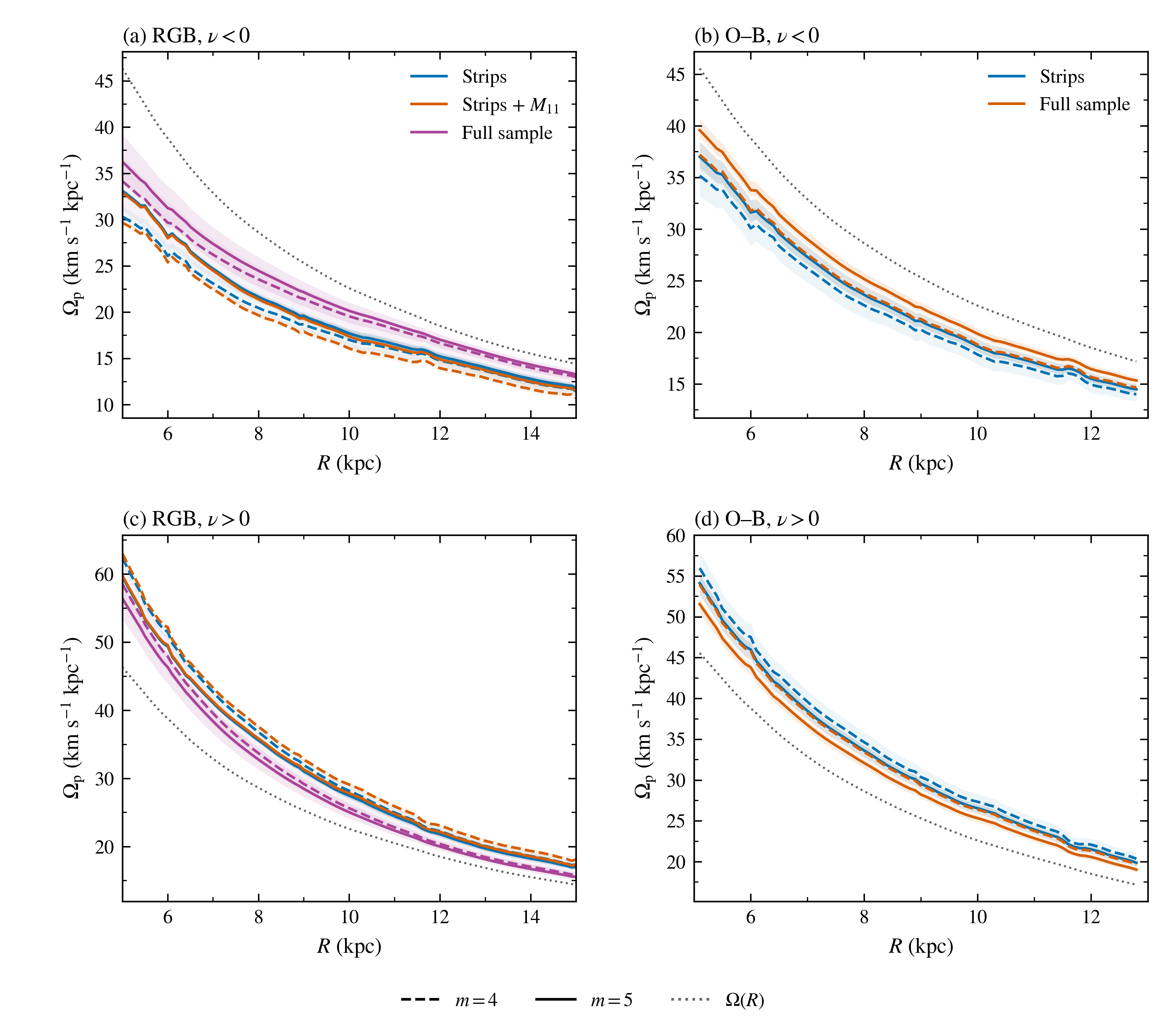}
\caption{Pattern-speed solutions derived from the radial velocity amplitude for $m=4$ (dashed curves) and $m=5$ (solid curves). Panels (a) and (c) show the RGB sample for $\nu<0$ and $\nu>0$, respectively, while panels (b) and (d) show the corresponding results for OB stars. Colours denote the sample selection. The shaded regions indicate the intervals between the 16th and 84th percentiles of the spatial block bootstrap distributions, and the gray dotted curve represents the angular velocity of Galactic circular rotation, $\Omega(R)$. For the OB sample, only the observed radial range is shown.}
\label{fig:pattern_speed_m4_m5}
\end{figure*}

For both stellar populations and both signs of the frequency, adopting $m = 5$ preserves the decrease in $\Omega_{\mathrm{p}}$ with increasing radius. For example, for the RGB stars in the stripes at $m = 5$, $\Omega_{\mathrm{p}}^{-}$ decreases from 33.11 to 11.80~\kmskpc\, between 5 and 15~kpc, while $\Omega_{\mathrm{p}}^{+}$ decreases from 59.60 to 17.02~\kmskpc. For the O--B stars in the stripes, over the range 5.1--12.8~kpc, the corresponding changes are 37.01 $\rightarrow$ 14.50 and 54.12 $\rightarrow$ 19.88~\kmskpc, respectively. Thus, adopting $m = 5$ does not result in a constant pattern speed.

\begin{table}
\centering
\caption{Phase differences between the radial and azimuthal velocity components for $m=4$ and $m=5$, in degrees. The values were obtained from the original data.}
\label{tab:phase_differences_m4_m5}
\setlength{\tabcolsep}{4pt}
\begin{tabular}{lcc}
\hline
Sample                   & $m=4$  & $m=5$  \\
\hline
RGB ``Strips''           & 132 [64,161] & 70 [31,111]  \\
RGB ``Strips $+M_{11}$'' & 116 [77,152] & 91 [51,136]  \\
RGB ``Full sample''      & 176 [51,163] & 104 [42,156] \\
OB ``Strips''            & 5 [6,56]     & 10 [5,53]  \\
OB ``Full sample''       & 12 [15,86]   & 6 [9,115]   \\
\hline
\end{tabular}

\end{table}

\section{Discussion and conclusions}
\label{sec:discussion}

We used 24,787 centroids within $5<R<15$~kpc and $131^\circ<\theta<269^\circ$, drawn from a catalogue constructed from 4,798,998 RGB stars. Using centroids reduces the contribution of random stellar velocities and enables a more reliable investigation of large-scale variations in the velocity field. The joint harmonic fit identifies periodic components in the centroid velocity field with amplitudes of a few \kms. Interpreting these components as the velocity response to a spiral wave, we determined the corresponding frequency parameters and inferred the pattern speeds at the radii of the spiral arms considered. We tested two predictions of the adopted model: the radial and azimuthal velocity components should exhibit a phase difference close to $90^\circ$, and the angular speed of a single rigidly rotating pattern should be constant.

Within the adopted single-mode Lin–Shu framework and the assumptions used to recover the kinematic response, our results do not simultaneously reproduce the expected phase quadrature and a radius-independent spiral pattern speed. The expected phase relation is not satisfied for either the spiral pattern or the individual arms. Furthermore, the observational data for RGB and O--B stars do not allow us to confirm a constant pattern speed for either the overall spiral pattern or the individual arms within the adopted Lin--Shu model. The main finding is that, when the amplitudes for each sample are determined under the approximation of weak radial variation, the inferred pattern speed generally decreases with Galactocentric radius. A similar radial dependence of $\Omega_{\mathrm{p}}(R)$ is obtained using only the radial amplitude and the perturbing potential adopted from \citet{Fernandez2008} and \citet{Eilers2020}. We regard this as our primary result because it does not depend on the phase relation between the two velocity components and does not use the less robustly determined azimuthal amplitude.

We also obtain decreasing profiles of $\Omega_{\mathrm{p}}(R)$ from the individual velocities of 32,358 O--B stars in $Gaia$ DR3, using the same calculation methods as for the RGB stars. Thus, the inferred radial dependence of $\Omega_{\mathrm{p}}(R)$ cannot be attributed solely to the large velocity dispersion of RGB stars. The calculations and comparisons between the individual arms and the overall pattern, between the RGB and O--B populations, and between the $m=4$ and $m=5$ models consistently indicate a radial dependence of $\Omega_{\mathrm{p}}(R)$, although they yield different numerical estimates of the pattern speed. We emphasise that, for both populations, the quantities $\nu(R)$ and $\Omega_{\mathrm{p}}(R)$ refer to the fitted harmonic component of the velocity field. Its geometric identification with the entire observed spiral pattern has not been established.

The observed velocity fields of the RGB centroids and O--B stars are therefore inconsistent with a description of the investigated region of the Galactic disk as a single rigidly rotating wave with the inferred pair of constant velocity-perturbation amplitudes. This conclusion is conditional on the adopted formulation and assumptions: the harmonic geometry, the rotation curve, the method used to separate the velocity components, the radius-independent amplitude ratio, the adopted velocity dispersion, and the selected solutions of the frequency equation.

Numerical experiments reported in the literature show that spiral structures can be transient, grow and fade, interact with a bar, or arise as the disk response to mass inhomogeneities. In such models, the angular speed of spiral structures need not be a single constant value; it may vary with radius and approach the local angular speed of the disk material. Our result is qualitatively consistent with this picture and with, for example, the $Gaia$ open-cluster results of \citet{Castro-Ginard2021} and the results of \citet{Liu2025}, who also found a generally decreasing pattern speed with radius, with most estimates lying close to the angular rotation curve. Similar behaviour occurs in models of transient, co-rotating spiral arms \citep{Grand2012a,Grand2012b,Roca-Fabrega2013}.

These comparisons do not establish that the structures share the same physical origin, but they show that the radial dependence of the pattern speed found here is not unique. In this sense, our results indicate that assigning a corotation radius to the entire investigated pattern is model-dependent and illustrate why a pattern speed measured near one radius should be extrapolated to a large region of the disk with caution \citep{Dias2019}. At the same time, these results provide quantitative constraints for comparisons between models and observations and suggest that the form of the radial dependence of the spiral pattern speed could serve as an additional dynamical criterion in searches for Milky Way analogues \citep{Vavilova2024}.

\section{Acknowledgements}
\label{sec:acknowledgements}

This work has made use of data from the European Space Agency (ESA) mission {\it Gaia} (\url{https://www.cosmos.esa.int/gaia}), processed by the {\it Gaia} Data Processing and Analysis Consortium (DPAC, \url{https://www.cosmos.esa.int/web/gaia/dpac/consortium}). Funding for the DPAC has been provided by national institutions, in particular the institutions participating in the {\it Gaia} Multilateral Agreement.

This work is supported by the National Research Foundation of Ukraine, project no. 2023.03/0188, the Ministry of Education
and Science of Ukraine, project no. 0126U001035 and the grant agreement $\Gamma$/19-25 between National Institute for Astrophysics (Italy) and V. N. Karazin Kharkiv National University (Ukraine).

\section*{Data availability}
\addcontentsline{toc}{section}{Data availability}
The used catalogue data is available in a standardised format for readers via the CDS (\url{https://cds.u-strasbg.fr}).
The software code used in this paper can be made available upon request by emailing the corresponding author.

\bibliographystyle{mnras}
\bibliography{references}

@ARTICLE{Akhmetov2024,
       author = {{Akhmetov}, V.~S. and {Bucciarelli}, B. and {Crosta}, M. and {Lattanzi}, M.~G. and {Spagna}, A. and {Re Fiorentin}, P. and {Bannikova}, E. Yu},
        title = "{A new kinematic model of the Galaxy: analysis of the stellar velocity field from Gaia Data Release 3}",
      journal = {\mnras},
         year = 2024,
        month = may,
       volume = {530},
       number = {1},
        pages = {710-729},
          doi = {10.1093/mnras/stae772},
archivePrefix = {arXiv},
       eprint = {2307.08527},
 primaryClass = {astro-ph.GA},
       adsurl = {https://ui.adsabs.harvard.edu/abs/2024MNRAS.530..710A}
}

@ARTICLE{Athanassoula2012,
       author = {{Athanassoula}, E.},
        title = "{Manifold-driven spirals in N-body barred galaxy simulations}",
      journal = {\mnras},
         year = 2012,
        month = oct,
       volume = {426},
       number = {1},
        pages = {L46-L50},
          doi = {10.1111/j.1745-3933.2012.01320.x},
archivePrefix = {arXiv},
       eprint = {1207.4590},
 primaryClass = {astro-ph.GA},
       adsurl = {https://ui.adsabs.harvard.edu/abs/2012MNRAS.426L..46A}
}

@ARTICLE{Baba2013,
       author = {{Baba}, Junichi and {Saitoh}, Takayuki R. and {Wada}, Keiichi},
        title = "{Dynamics of Non-steady Spiral Arms in Disk Galaxies}",
      journal = {\apj},
         year = 2013,
        month = jan,
       volume = {763},
       number = {1},
          eid = {46},
        pages = {46},
          doi = {10.1088/0004-637X/763/1/46},
archivePrefix = {arXiv},
       eprint = {1211.5401},
 primaryClass = {astro-ph.GA},
       adsurl = {https://ui.adsabs.harvard.edu/abs/2013ApJ...763...46B}
}

@ARTICLE{Bertin1989,
       author = {{Bertin}, G. and {Lin}, C.~C. and {Lowe}, S.~A. and {Thurstans}, R.~P.},
        title = "{Modal Approach to the Morphology of Spiral Galaxies. II. Dynamical Mechanisms}",
      journal = {\apj},
         year = 1989,
        month = mar,
       volume = {338},
        pages = {104},
          doi = {10.1086/167183},
       adsurl = {https://ui.adsabs.harvard.edu/abs/1989ApJ...338..104B}
}

@BOOK{Bertin1996,
       author = {{Bertin}, G. and {Lin}, C.~C.},
        title = "{Spiral structure in galaxies a density wave theory}",
         year = 1996,
       adsurl = {https://ui.adsabs.harvard.edu/abs/1996ssgd.book.....B},
      publisher = {Cambridge, MA MIT Press}
}

@ARTICLE{Bobylev2025,
       author = {{Bobylev}, V.~V. and {Bajkova}, A.~T. and {Smirnov}, A.~A.},
        title = "{The Spiral Pattern Speed in the Milky Way Galaxy}",
      journal = {Astronomy Letters},
         year = 2025,
        month = may,
       volume = {51},
       number = {5},
        pages = {278-286},
          doi = {10.1134/S1063773725700422},
archivePrefix = {arXiv},
       eprint = {2510.26450},
 primaryClass = {astro-ph.GA},
       adsurl = {https://ui.adsabs.harvard.edu/abs/2025AstL...51..278B}
}

@ARTICLE{Castro-Ginard2021,
       author = {{Castro-Ginard}, A. and {McMillan}, P.~J. and {Luri}, X. and {Jordi}, C. and {Romero-G{\'o}mez}, M. and {Cantat-Gaudin}, T. and {Casamiquela}, L. and {Tarricq}, Y. and {Soubiran}, C. and {Anders}, F.},
        title = "{Milky Way spiral arms from open clusters in Gaia EDR3}",
      journal = {\aap},
         year = 2021,
        month = aug,
       volume = {652},
          eid = {A162},
        pages = {A162},
          doi = {10.1051/0004-6361/202039751},
archivePrefix = {arXiv},
       eprint = {2105.04590},
 primaryClass = {astro-ph.GA},
       adsurl = {https://ui.adsabs.harvard.edu/abs/2021A&A...652A.162C}
}

@ARTICLE{Churchwell2009,
       author = {{Churchwell}, Ed and {Babler}, Brian L. and {Meade}, Marilyn R. and {Whitney}, Barbara A. and {Benjamin}, Robert and {Indebetouw}, Remy and {Cyganowski}, Claudia and {Robitaille}, Thomas P. and {Povich}, Matthew and {Watson}, Christer and {Bracker}, Steve},
        title = "{The Spitzer/GLIMPSE Surveys: A New View of the Milky Way}",
      journal = {\pasp},
         year = 2009,
        month = mar,
       volume = {121},
       number = {877},
        pages = {213},
          doi = {10.1086/597811},
       adsurl = {https://ui.adsabs.harvard.edu/abs/2009PASP..121..213C}
}

@ARTICLE{Denyshchenko2024,
       author = {{Denyshchenko}, S.~I. and {Fedorov}, P.~N. and {Akhmetov}, V.~S. and {Velichko}, A.~B. and {Dmytrenko}, A.~M.},
        title = "{Determining the parameters of the spiral arms of the Galaxy from kinematic tracers based on Gaia DR3 data}",
      journal = {\mnras},
         year = 2024,
        month = jan,
       volume = {527},
       number = {1},
        pages = {1472-1480},
          doi = {10.1093/mnras/stad3350},
archivePrefix = {arXiv},
       eprint = {2308.14021},
 primaryClass = {astro-ph.GA},
       adsurl = {https://ui.adsabs.harvard.edu/abs/2024MNRAS.527.1472D}
}

@ARTICLE{Dias2019,
       author = {{Dias}, W.~S. and {Monteiro}, H. and {L{\'e}pine}, J.~R.~D. and {Barros}, D.~A.},
        title = "{The spiral pattern rotation speed of the Galaxy and the corotation radius with Gaia DR2}",
      journal = {\mnras},
         year = 2019,
        month = jul,
       volume = {486},
       number = {4},
        pages = {5726-5736},
          doi = {10.1093/mnras/stz1196},
archivePrefix = {arXiv},
       eprint = {1905.08133},
 primaryClass = {astro-ph.GA},
       adsurl = {https://ui.adsabs.harvard.edu/abs/2019MNRAS.486.5726D}
}

@ARTICLE{Dobbs2010,
       author = {{Dobbs}, C.~L. and {Theis}, C. and {Pringle}, J.~E. and {Bate}, M.~R.},
        title = "{Simulations of the grand design galaxy M51: a case study for analysing tidally induced spiral structure}",
      journal = {\mnras},
         year = 2010,
        month = apr,
       volume = {403},
       number = {2},
        pages = {625-645},
          doi = {10.1111/j.1365-2966.2009.16161.x},
archivePrefix = {arXiv},
       eprint = {0912.1201},
 primaryClass = {astro-ph.GA},
       adsurl = {https://ui.adsabs.harvard.edu/abs/2010MNRAS.403..625D}
}

@ARTICLE{Drimmel2025,
       author = {{Drimmel}, R. and {Khanna}, S. and {Poggio}, E. and {Skowron}, D.~M.},
        title = "{The Milky Way as seen by classical Cepheids: II. Spiral structure}",
      journal = {\aap},
         year = 2025,
        month = jun,
       volume = {698},
          eid = {A230},
        pages = {A230},
          doi = {10.1051/0004-6361/202451100},
archivePrefix = {arXiv},
       eprint = {2406.09127},
 primaryClass = {astro-ph.GA},
       adsurl = {https://ui.adsabs.harvard.edu/abs/2025A&A...698A.230D}
}

@ARTICLE{Eilers2020,
       author = {{Eilers}, Anna-Christina and {Hogg}, David W. and {Rix}, Hans-Walter and {Frankel}, Neige and {Hunt}, Jason A.~S. and {Fouvry}, Jean-Baptiste and {Buck}, Tobias},
        title = "{The Strength of the Dynamical Spiral Perturbation in the Galactic Disk}",
      journal = {\apj},
         year = 2020,
        month = sep,
       volume = {900},
       number = {2},
          eid = {186},
        pages = {186},
          doi = {10.3847/1538-4357/abac0b},
archivePrefix = {arXiv},
       eprint = {2003.01132},
 primaryClass = {astro-ph.GA},
       adsurl = {https://ui.adsabs.harvard.edu/abs/2020ApJ...900..186E}
}

@ARTICLE{Elmegreen1982,
       author = {{Elmegreen}, D.~M. and {Elmegreen}, B.~G.},
        title = "{Flocculent and grand design spiral structure in field, binary and group galaxies.}",
      journal = {\mnras},
         year = 1982,
        month = dec,
       volume = {201},
        pages = {1021-1034},
          doi = {10.1093/mnras/201.4.1021},
       adsurl = {https://ui.adsabs.harvard.edu/abs/1982MNRAS.201.1021E}
}

@ARTICLE{Fedorov2023,
       author = {{Fedorov}, P.~N. and {Akhmetov}, V.~S. and {Velichko}, A.~B. and {Dmytrenko}, A.~M. and {Denyshchenko}, S.~I.},
        title = "{Mapping the kinematic parameters of the Galaxy from the Gaia EDR3 red giants and sub-giants}",
      journal = {\mnras},
         year = 2023,
        month = jan,
       volume = {518},
       number = {2},
        pages = {2761-2774},
          doi = {10.1093/mnras/stac3218},
archivePrefix = {arXiv},
       eprint = {2205.09381},
 primaryClass = {astro-ph.GA},
       adsurl = {https://ui.adsabs.harvard.edu/abs/2023MNRAS.518.2761F}
}

@ARTICLE{Fedorov2026,
       author = {{Fedorov}, P.~N. and {Dmytrenko}, A.~M. and {Akhmetov}, V.~S. and {Velichko}, A.~B. and {Zheleznyak}, A.~P. and {Khramtsov}, V.~P. and {Denyshchenko}, S.~I. and {Vavilova}, I.~B. and {Dobrycheva}, D.~V. and {Sergijenko}, O. and {Vasylenko}, A.~A. and {Kompaniiets}, O.~V.},
        title = "{Local Jeans estimates of the Galaxy's circular velocity, based on RGB stars from the Gaia catalogue}",
      journal = {\mnras},
         year = 2026,
        month = aug,
       volume = {550},
       number = {4},
          eid = {stag1310},
        pages = {stag1310},
          doi = {10.1093/mnras/stag1310},
archivePrefix = {arXiv},
       eprint = {2511.22295},
 primaryClass = {astro-ph.GA},
       adsurl = {https://ui.adsabs.harvard.edu/abs/2026MNRAS.550g1310F}
}

@ARTICLE{Garma-Oehmichen2021,
       author = {{Garma-Oehmichen}, L. and {Martinez-Medina}, L. and {Hern{\'a}ndez-Toledo}, H. and {Puerari}, I.},
        title = "{How the bar properties affect the induced spiral structure}",
      journal = {\mnras},
         year = 2021,
        month = apr,
       volume = {502},
       number = {4},
        pages = {4708-4722},
          doi = {10.1093/mnras/stab333},
archivePrefix = {arXiv},
       eprint = {2102.01134},
 primaryClass = {astro-ph.GA},
       adsurl = {https://ui.adsabs.harvard.edu/abs/2021MNRAS.502.4708G}
}

@ARTICLE{Goldreich1965,
       author = {{Goldreich}, P. and {Lynden-Bell}, D.},
        title = "{II. Spiral arms as sheared gravitational instabilities}",
      journal = {\mnras},
         year = 1965,
        month = jan,
       volume = {130},
        pages = {125},
          doi = {10.1093/mnras/130.2.125},
       adsurl = {https://ui.adsabs.harvard.edu/abs/1965MNRAS.130..125G}
}

@ARTICLE{Grand2012a,
       author = {{Grand}, Robert J.~J. and {Kawata}, Daisuke and {Cropper}, Mark},
        title = "{The dynamics of stars around spiral arms}",
      journal = {\mnras},
         year = 2012,
        month = apr,
       volume = {421},
       number = {2},
        pages = {1529-1538},
          doi = {10.1111/j.1365-2966.2012.20411.x},
archivePrefix = {arXiv},
       eprint = {1112.0019},
 primaryClass = {astro-ph.GA},
       adsurl = {https://ui.adsabs.harvard.edu/abs/2012MNRAS.421.1529G}
}

@ARTICLE{Grand2012b,
       author = {{Grand}, Robert J.~J. and {Kawata}, Daisuke and {Cropper}, Mark},
        title = "{Dynamics of stars around spiral arms in an N-body/SPH simulated barred spiral galaxy}",
      journal = {\mnras},
         year = 2012,
        month = oct,
       volume = {426},
       number = {1},
        pages = {167-180},
          doi = {10.1111/j.1365-2966.2012.21733.x},
archivePrefix = {arXiv},
       eprint = {1202.6387},
 primaryClass = {astro-ph.GA},
       adsurl = {https://ui.adsabs.harvard.edu/abs/2012MNRAS.426..167G}
}

@ARTICLE{GRAVITY2021,
       author = {{GRAVITY Collaboration} and {Abuter}, R. and {Amorim}, A. and {Baub{\"o}ck}, M. and {Berger}, J.~P. and {Bonnet}, H. and {Brandner}, W. and {Cl{\'e}net}, Y. and {Davies}, R. and {de Zeeuw}, P.~T. and {Dexter}, J. and {Dallilar}, Y. and {Drescher}, A. and {Eckart}, A. and {Eisenhauer}, F. and {F{\"o}rster Schreiber}, N.~M. and {Garcia}, P. and {Gao}, F. and {Gendron}, E. and {Genzel}, R. and {Gillessen}, S. and {Habibi}, M. and {Haubois}, X. and {Hei{\ss}el}, G. and {Henning}, T. and {Hippler}, S. and {Horrobin}, M. and {Jim{\'e}nez-Rosales}, A. and {Jochum}, L. and {Jocou}, L. and {Kaufer}, A. and {Kervella}, P. and {Lacour}, S. and {Lapeyr{\`e}re}, V. and {Le Bouquin}, J.-B. and {L{\'e}na}, P. and {Lutz}, D. and {Nowak}, M. and {Ott}, T. and {Paumard}, T. and {Perraut}, K. and {Perrin}, G. and {Pfuhl}, O. and {Rabien}, S. and {Rodr{\'\i}guez-Coira}, G. and {Shangguan}, J. and {Shimizu}, T. and {Scheithauer}, S. and {Stadler}, J. and {Straub}, O. and {Straubmeier}, C. and {Sturm}, E. and {Tacconi}, L.~J. and {Vincent}, F. and {von Fellenberg}, S. and {Waisberg}, I. and {Widmann}, F. and {Wieprecht}, E. and {Wiezorrek}, E. and {Woillez}, J. and {Yazici}, S. and {Young}, A. and {Zins}, G.},
        title = "{Improved GRAVITY astrometric accuracy from modeling optical aberrations}",
      journal = {\aap},
         year = 2021,
        month = mar,
       volume = {647},
          eid = {A59},
        pages = {A59},
          doi = {10.1051/0004-6361/202040208},
archivePrefix = {arXiv},
       eprint = {2101.12098},
 primaryClass = {astro-ph.GA},
       adsurl = {https://ui.adsabs.harvard.edu/abs/2021A&A...647A..59G}
}

@ARTICLE{Griv2026,
       author = {{Griv}, Evgeny and {Akhmetov}, Volodymyr and {Yahalom}, Asher},
        title = "{The spiral pattern rotation speed of the Milky Way Galaxy in the Lin-Shu theory of small-amplitude density waves}",
      journal = {\apss},
         year = 2026,
        month = jun,
       volume = {371},
       number = {6},
          eid = {64},
        pages = {64},
          doi = {10.1007/s10509-026-04594-0},
       adsurl = {https://ui.adsabs.harvard.edu/abs/2026Ap&SS.371...64G}
}

@ARTICLE{Joshi2023,
       author = {{Joshi}, Yogesh C. and {Malhotra}, Sagar},
        title = "{Revisiting Galactic Disk and Spiral Arms Using Open Clusters}",
      journal = {\aj},
         year = 2023,
        month = oct,
       volume = {166},
       number = {4},
          eid = {170},
        pages = {170},
          doi = {10.3847/1538-3881/acf7c8},
archivePrefix = {arXiv},
       eprint = {2212.09384},
 primaryClass = {astro-ph.GA},
       adsurl = {https://ui.adsabs.harvard.edu/abs/2023AJ....166..170J}
}

@ARTICLE{Julian1966,
       author = {{Julian}, William H. and {Toomre}, Alar},
        title = "{Non-Axisymmetric Responses of Differentially Rotating Disks of Stars}",
      journal = {\apj},
         year = 1966,
        month = dec,
       volume = {146},
        pages = {810},
          doi = {10.1086/148957},
       adsurl = {https://ui.adsabs.harvard.edu/abs/1966ApJ...146..810J}
}

@ARTICLE{Kendall2011,
       author = {{Kendall}, S. and {Kennicutt}, R.~C. and {Clarke}, C.},
        title = "{Spiral structure in nearby galaxies - I. Sample, data analysis and overview of results}",
      journal = {\mnras},
         year = 2011,
        month = jun,
       volume = {414},
       number = {1},
        pages = {538-564},
          doi = {10.1111/j.1365-2966.2011.18422.x},
archivePrefix = {arXiv},
       eprint = {1101.5764},
 primaryClass = {astro-ph.GA},
       adsurl = {https://ui.adsabs.harvard.edu/abs/2011MNRAS.414..538K}
}

@ARTICLE{Khanna2025,
       author = {{Khanna}, Shourya and {Yu}, Jie and {Drimmel}, Ronald and {Poggio}, Eloisa and {Cantat-Gaudin}, Tristan and {Castro-Ginard}, Alfred and {Kurbatov}, Evgeny and {Belokurov}, Vasily and {Brown}, Anthony and {Fouesneau}, Morgan and {Casey}, Andrew and {Rix}, Hans-Walter},
        title = "{GaiaUnlimited: The old stellar disc of the Milky Way as traced by the red clump}",
      journal = {\aap},
         year = 2025,
        month = sep,
       volume = {701},
          eid = {A270},
        pages = {A270},
          doi = {10.1051/0004-6361/202452798},
archivePrefix = {arXiv},
       eprint = {2410.22036},
 primaryClass = {astro-ph.GA},
       adsurl = {https://ui.adsabs.harvard.edu/abs/2025A&A...701A.270K}
}

@ARTICLE{Kormendy1979,
       author = {{Kormendy}, J. and {Norman}, C.~A.},
        title = "{Observational constraints on driving mechanisms for spiral density waves.}",
      journal = {\apj},
         year = 1979,
        month = oct,
       volume = {233},
        pages = {539-552},
          doi = {10.1086/157414},
       adsurl = {https://ui.adsabs.harvard.edu/abs/1979ApJ...233..539K}
}

@ARTICLE{Lin1969,
       author = {{Lin}, C.~C. and {Yuan}, C. and {Shu}, Frank H.},
        title = "{On the Spiral Structure of Disk Galaxies. III. Comparison with Observations}",
      journal = {\apj},
         year = 1969,
        month = mar,
       volume = {155},
        pages = {721},
          doi = {10.1086/149907},
       adsurl = {https://ui.adsabs.harvard.edu/abs/1969ApJ...155..721L}
}

@ARTICLE{Lin1964,
       author = {{Lin}, C.~C. and {Shu}, Frank H.},
        title = "{On the Spiral Structure of Disk Galaxies.}",
      journal = {\apj},
         year = 1964,
        month = aug,
       volume = {140},
        pages = {646},
          doi = {10.1086/147955},
       adsurl = {https://ui.adsabs.harvard.edu/abs/1964ApJ...140..646L}
}

@ARTICLE{DOnghia2013,
       author = {{D'Onghia}, Elena and {Vogelsberger}, Mark and {Hernquist}, Lars},
        title = "{Self-perpetuating Spiral Arms in Disk Galaxies}",
      journal = {\apj},
         year = 2013,
        month = mar,
       volume = {766},
       number = {1},
          eid = {34},
        pages = {34},
          doi = {10.1088/0004-637X/766/1/34},
archivePrefix = {arXiv},
       eprint = {1204.0513},
 primaryClass = {astro-ph.GA},
       adsurl = {https://ui.adsabs.harvard.edu/abs/2013ApJ...766...34D}
}

@ARTICLE{Pettitt2017,
       author = {{Pettitt}, Alex R. and {Tasker}, Elizabeth J. and {Wadsley}, James W. and {Keller}, Ben W. and {Benincasa}, Samantha M.},
        title = "{Star formation and ISM morphology in tidally induced spiral structures}",
      journal = {\mnras},
         year = 2017,
        month = jul,
       volume = {468},
       number = {4},
        pages = {4189-4204},
          doi = {10.1093/mnras/stx736},
archivePrefix = {arXiv},
       eprint = {1704.04127},
 primaryClass = {astro-ph.GA},
       adsurl = {https://ui.adsabs.harvard.edu/abs/2017MNRAS.468.4189P}
}

@ARTICLE{Poggio2021,
       author = {{Poggio}, E. and {Drimmel}, R. and {Cantat-Gaudin}, T. and {Ramos}, P. and {Ripepi}, V. and {Zari}, E. and {Andrae}, R. and {Blomme}, R. and {Chemin}, L. and {Clementini}, G. and {Figueras}, F. and {Fouesneau}, M. and {Fr{\'e}mat}, Y. and {Lobel}, A. and {Marshall}, D.~J. and {Muraveva}, T. and {Romero-G{\'o}mez}, M.},
        title = "{Galactic spiral structure revealed by Gaia EDR3}",
      journal = {\aap},
         year = 2021,
        month = jul,
       volume = {651},
          eid = {A104},
        pages = {A104},
          doi = {10.1051/0004-6361/202140687},
archivePrefix = {arXiv},
       eprint = {2103.01970},
 primaryClass = {astro-ph.GA},
       adsurl = {https://ui.adsabs.harvard.edu/abs/2021A&A...651A.104P}
}

@ARTICLE{Purcell2011,
       author = {{Purcell}, Chris W. and {Bullock}, James S. and {Tollerud}, Erik J. and {Rocha}, Miguel and {Chakrabarti}, Sukanya},
        title = "{The Sagittarius impact as an architect of spirality and outer rings in the Milky Way}",
      journal = {\nat},
         year = 2011,
        month = sep,
       volume = {477},
       number = {7364},
        pages = {301-303},
          doi = {10.1038/nature10417},
archivePrefix = {arXiv},
       eprint = {1109.2918},
 primaryClass = {astro-ph.GA},
       adsurl = {https://ui.adsabs.harvard.edu/abs/2011Natur.477..301P}
}

@ARTICLE{Reid2004,
       author = {{Reid}, M.~J. and {Brunthaler}, A.},
        title = "{The Proper Motion of Sagittarius A*. II. The Mass of Sagittarius A*}",
      journal = {\apj},
         year = 2004,
        month = dec,
       volume = {616},
       number = {2},
        pages = {872-884},
          doi = {10.1086/424960},
archivePrefix = {arXiv},
       eprint = {astro-ph/0408107},
 primaryClass = {astro-ph},
       adsurl = {https://ui.adsabs.harvard.edu/abs/2004ApJ...616..872R}
}

@INCOLLECTION{Rohlfs1977a,
       author = {{Rohlfs}, Kristen},
        title = "{The theory of spiral structure of galaxies}",
    booktitle = {Lectures on Density Wave Theory},
         year = 1977,
       editor = {{Rohlfs}, Kristen},
       volume = {69},
        pages = {64-105},
          doi = {10.1007/3540084487_5},
       adsurl = {https://ui.adsabs.harvard.edu/abs/1977LNP....69...64R}
}

@ARTICLE{Roca-Fabrega2013,
       author = {{Roca-F{\`a}brega}, Santi and {Valenzuela}, Octavio and {Figueras}, Francesca and {Romero-G{\'o}mez}, Merc{\`e} and {Vel{\'a}zquez}, H{\'e}ctor and {Antoja}, Teresa and {Pichardo}, B{\'a}rbara},
        title = "{On galaxy spiral arms' nature as revealed by rotation frequencies}",
      journal = {\mnras},
         year = 2013,
        month = jul,
       volume = {432},
       number = {4},
        pages = {2878-2885},
          doi = {10.1093/mnras/stt643},
archivePrefix = {arXiv},
       eprint = {1302.6981},
 primaryClass = {astro-ph.GA},
       adsurl = {https://ui.adsabs.harvard.edu/abs/2013MNRAS.432.2878R}
}

@ARTICLE{Salo2000a,
       author = {{Salo}, Heikki and {Laurikainen}, Eija},
        title = "{N-body model for M51 - I. Multiple encounter versus single passage?}",
      journal = {\mnras},
         year = 2000,
        month = dec,
       volume = {319},
       number = {2},
        pages = {377-392},
          doi = {10.1046/j.1365-8711.2000.03650.x},
       adsurl = {https://ui.adsabs.harvard.edu/abs/2000MNRAS.319..377S}
}

@ARTICLE{Salo2000b,
       author = {{Salo}, Heikki and {Laurikainen}, Eija},
        title = "{N-body model for M51 - II. Inner structure}",
      journal = {\mnras},
         year = 2000,
        month = dec,
       volume = {319},
       number = {2},
        pages = {393-413},
          doi = {10.1046/j.1365-8711.2000.03651.x},
       adsurl = {https://ui.adsabs.harvard.edu/abs/2000MNRAS.319..393S}
}

@ARTICLE{Sellwood2022,
       author = {{Sellwood}, J.~A. and {Carlberg}, R.~G.},
        title = "{Spiral instabilities: mode saturation and decay}",
      journal = {\mnras},
         year = 2022,
        month = dec,
       volume = {517},
       number = {2},
        pages = {2610-2621},
          doi = {10.1093/mnras/stac2817},
archivePrefix = {arXiv},
       eprint = {2209.14256},
 primaryClass = {astro-ph.GA},
       adsurl = {https://ui.adsabs.harvard.edu/abs/2022MNRAS.517.2610S}
}

@ARTICLE{Sellwood2019,
       author = {{Sellwood}, J.~A. and {Carlberg}, Ray G.},
        title = "{Spiral instabilities: mechanism for recurrence}",
      journal = {\mnras},
         year = 2019,
        month = oct,
       volume = {489},
       number = {1},
        pages = {116-131},
          doi = {10.1093/mnras/stz2132},
archivePrefix = {arXiv},
       eprint = {1906.04191},
 primaryClass = {astro-ph.GA},
       adsurl = {https://ui.adsabs.harvard.edu/abs/2019MNRAS.489..116S}
}

@ARTICLE{Sellwood2014,
       author = {{Sellwood}, J.~A. and {Carlberg}, R.~G.},
        title = "{Transient Spirals as Superposed Instabilities}",
      journal = {\apj},
         year = 2014,
        month = apr,
       volume = {785},
       number = {2},
          eid = {137},
        pages = {137},
          doi = {10.1088/0004-637X/785/2/137},
archivePrefix = {arXiv},
       eprint = {1403.1135},
 primaryClass = {astro-ph.GA},
       adsurl = {https://ui.adsabs.harvard.edu/abs/2014ApJ...785..137S}
}

@ARTICLE{Tremaine1984,
       author = {{Tremaine}, S. and {Weinberg}, M.~D.},
        title = "{A kinematic method for measuring the pattern speed of barred galaxies.}",
      journal = {\apjl},
         year = 1984,
        month = jul,
       volume = {282},
        pages = {L5-L7},
          doi = {10.1086/184292},
       adsurl = {https://ui.adsabs.harvard.edu/abs/1984ApJ...282L...5T}
}

@ARTICLE{Vallenari2023,
       author = {{Gaia Collaboration} and {Vallenari}, A. and {Brown}, A.~G.~A. and {Prusti}, T. and {de Bruijne}, J.~H.~J. and {Arenou}, F. and {Babusiaux}, C. and {Biermann}, M. and {Creevey}, O.~L. and {Ducourant}, C. and {Evans}, D.~W. and {Eyer}, L. and {Guerra}, R. and {Hutton}, A. and {Jordi}, C. and {Klioner}, S.~A. and {Lammers}, U.~L. and {Lindegren}, L. and {Luri}, X. and {Mignard}, F. and {Panem}, C. and {Pourbaix}, D. and {Randich}, S. and {Sartoretti}, P. and {Soubiran}, C. and {Tanga}, P. and {Walton}, N.~A. and {Bailer-Jones}, C.~A.~L. and {Bastian}, U. and {Drimmel}, R. and {Jansen}, F. and {Katz}, D. and {Lattanzi}, M.~G. and {van Leeuwen}, F. and {Bakker}, J. and {Cacciari}, C. and {Casta{\~n}eda}, J. and {De Angeli}, F. and {Fabricius}, C. and {Fouesneau}, M. and {Fr{\'e}mat}, Y. and {Galluccio}, L. and {Guerrier}, A. and {Heiter}, U. and {Masana}, E. and {Messineo}, R. and {Mowlavi}, N. and {Nicolas}, C. and {Nienartowicz}, K. and {Pailler}, F. and {Panuzzo}, P. and {Riclet}, F. and {Roux}, W. and {Seabroke}, G.~M. and {Sordo}, R. and {Th{\'e}venin}, F. and {Gracia-Abril}, G. and {Portell}, J. and {Teyssier}, D. and {Altmann}, M. and {Andrae}, R. and {Audard}, M. and {Bellas-Velidis}, I. and {Benson}, K. and {Berthier}, J. and {Blomme}, R. and {Burgess}, P.~W. and {Busonero}, D. and {Busso}, G. and {C{\'a}novas}, H. and {Carry}, B. and {Cellino}, A. and {Cheek}, N. and {Clementini}, G. and {Damerdji}, Y. and {Davidson}, M. and {de Teodoro}, P. and {Nu{\~n}ez Campos}, M. and {Delchambre}, L. and {Dell'Oro}, A. and {Esquej}, P. and {Fern{\'a}ndez-Hern{\'a}ndez}, J. and {Fraile}, E. and {Garabato}, D. and {Garc{\'\i}a-Lario}, P. and {Gosset}, E. and {Haigron}, R. and {Halbwachs}, J.-L. and {Hambly}, N.~C. and {Harrison}, D.~L. and {Hern{\'a}ndez}, J. and {Hestroffer}, D. and {Hodgkin}, S.~T. and {Holl}, B. and {Jan{\ss}en}, K. and {Jevardat de Fombelle}, G. and {Jordan}, S. and {Krone-Martins}, A. and {Lanzafame}, A.~C. and {L{\"o}ffler}, W. and {Marchal}, O. and {Marrese}, P.~M. and {Moitinho}, A. and {Muinonen}, K. and {Osborne}, P. and {Pancino}, E. and {Pauwels}, T. and {Recio-Blanco}, A. and {Reyl{\'e}}, C. and {Riello}, M. and {Rimoldini}, L. and {Roegiers}, T. and {Rybizki}, J. and {Sarro}, L.~M. and {Siopis}, C. and {Smith}, M. and {Sozzetti}, A. and {Utrilla}, E. and {van Leeuwen}, M. and {Abbas}, U. and {{\'A}brah{\'a}m}, P. and {Abreu Aramburu}, A. and {Aerts}, C. and {Aguado}, J.~J. and {Ajaj}, M. and {Aldea-Montero}, F. and {Altavilla}, G. and {{\'A}lvarez}, M.~A. and {Alves}, J. and {Anders}, F. and {Anderson}, R.~I. and {Anglada Varela}, E. and {Antoja}, T. and {Baines}, D. and {Baker}, S.~G. and {Balaguer-N{\'u}{\~n}ez}, L. and {Balbinot}, E. and {Balog}, Z. and {Barache}, C. and {Barbato}, D. and {Barros}, M. and {Barstow}, M.~A. and {Bartolom{\'e}}, S. and {Bassilana}, J.-L. and {Bauchet}, N. and {Becciani}, U. and {Bellazzini}, M. and {Berihuete}, A. and {Bernet}, M. and {Bertone}, S. and {Bianchi}, L. and {Binnenfeld}, A. and {Blanco-Cuaresma}, S. and {Blazere}, A. and {Boch}, T. and {Bombrun}, A. and {Bossini}, D. and {Bouquillon}, S. and {Bragaglia}, A. and {Bramante}, L. and {Breedt}, E. and {Bressan}, A. and {Brouillet}, N. and {Brugaletta}, E. and {Bucciarelli}, B. and {Burlacu}, A. and {Butkevich}, A.~G. and {Buzzi}, R. and {Caffau}, E. and {Cancelliere}, R. and {Cantat-Gaudin}, T. and {Carballo}, R. and {Carlucci}, T. and {Carnerero}, M.~I. and {Carrasco}, J.~M. and {Casamiquela}, L. and {Castellani}, M. and {Castro-Ginard}, A. and {Chaoul}, L. and {Charlot}, P. and {Chemin}, L. and {Chiaramida}, V. and {Chiavassa}, A. and {Chornay}, N. and {Comoretto}, G. and {Contursi}, G. and {Cooper}, W.~J. and {Cornez}, T. and {Cowell}, S. and {Crifo}, F. and {Cropper}, M. and {Crosta}, M. and {Crowley}, C. and {Dafonte}, C. and {Dapergolas}, A. and {David}, M. and {David}, P. and {de Laverny}, P. and {De Luise}, F. and {De March}, R.},
        title = "{Gaia Data Release 3. Summary of the content and survey properties}",
      journal = {\aap},
         year = 2023,
        month = jun,
       volume = {674},
          eid = {A1},
        pages = {A1},
          doi = {10.1051/0004-6361/202243940},
archivePrefix = {arXiv},
       eprint = {2208.00211},
 primaryClass = {astro-ph.GA},
       adsurl = {https://ui.adsabs.harvard.edu/abs/2023A&A...674A...1G}
}

@ARTICLE{Siebert2012,
       author = {{Siebert}, A. and {Famaey}, B. and {Binney}, J. and {Burnett}, B. and {Faure}, C. and {Minchev}, I. and {Williams}, M.~E.~K. and {Bienaym{\'e}}, O. and {Bland-Hawthorn}, J. and {Boeche}, C. and {Gibson}, B.~K. and {Grebel}, E.~K. and {Helmi}, A. and {Just}, A. and {Munari}, U. and {Navarro}, J.~F. and {Parker}, Q.~A. and {Reid}, W.~A. and {Seabroke}, G. and {Siviero}, A. and {Steinmetz}, M. and {Zwitter}, T.},
        title = "{The properties of the local spiral arms from RAVE data: two-dimensional density wave approach}",
      journal = {\mnras},
         year = 2012,
        month = sep,
       volume = {425},
       number = {3},
        pages = {2335-2342},
          doi = {10.1111/j.1365-2966.2012.21638.x},
archivePrefix = {arXiv},
       eprint = {1207.0363},
 primaryClass = {astro-ph.GA},
       adsurl = {https://ui.adsabs.harvard.edu/abs/2012MNRAS.425.2335S}
}

@ARTICLE{Fernandez2008,
       author = {{Fern{\'a}ndez}, D. and {Figueras}, F. and {Torra}, J.},
        title = "{On the kinematic evolution of young local associations and the Scorpius-Centaurus complex}",
      journal = {\aap},
         year = 2008,
        month = mar,
       volume = {480},
       number = {3},
        pages = {735-751},
          doi = {10.1051/0004-6361:20077720},
archivePrefix = {arXiv},
       eprint = {0801.0605},
 primaryClass = {astro-ph},
       adsurl = {https://ui.adsabs.harvard.edu/abs/2008A&A...480..735F}
}

@ARTICLE{Liu2025,
       author = {{Liu}, Xiaochen and {He}, Zhihong and {Luo}, Yangping and {Wang}, Kun},
        title = "{The disrupting and growing open cluster spiral arm patterns of the Milky Way}",
      journal = {\mnras},
         year = 2025,
        month = mar,
       volume = {537},
       number = {3},
        pages = {2403-2411},
          doi = {10.1093/mnras/staf153},
archivePrefix = {arXiv},
       eprint = {2501.14215},
 primaryClass = {astro-ph.GA},
       adsurl = {https://ui.adsabs.harvard.edu/abs/2025MNRAS.537.2403L}
}

@ARTICLE{Vavilova2024,
       author = {{Vavilova}, I.~B. and {Fedorov}, P.~M. and {Dobrycheva}, D.~V. and {Sergijenko}, O. and {Vasylenko}, A.~A. and {Dmytrenko}, A.~M. and {Khramtsov}, V.~P. and {Kompaniiets}, O.~V.},
        title = "{An advanced approach to the definition of the ``Milky Way galaxies-analogues''}",
      journal = {Kosmichna Nauka i Tekhnologiya},
         year = 2024,
        month = jan,
       volume = {30},
       number = {4},
        pages = {81},
          doi = {10.15407/knit2024.04.081},
archivePrefix = {arXiv},
       eprint = {2410.04411},
 primaryClass = {astro-ph.GA},
       adsurl = {https://ui.adsabs.harvard.edu/abs/2024KosNT..30d..81V}
}

@ARTICLE{Vavilova2026a,
    author = {{Vavilova}, I.~B. and {Dmytrenko}, A.~M. and {Dobrycheva}, D.~V. and {Fedorov}, P.~N. and {Izviekova}, I.~O. and {Khramtsov}, V.~P. and {Kompaniiets}, O.~V. and {Kukhar}, O.~N. and {Pastoven}, O.~S. and {Sergijenko}, O. and {Vasylenko}, A.~A.},
    title = "{Milky Way Near Twins (MWNeTs). I. A Hierarchical Framework for Identifying Candidate Evolutionary Counterparts of the Milky Way}",
    journal = {arXiv e-prints},
    year = 2026,
    month = jul,
    eid = {arXiv:2607.08336},
    pages = {arXiv:2607.08336},
    doi = {10.48550/arXiv.2607.08336},
    archivePrefix = {arXiv},
    eprint = {2607.08336},
    primaryClass = {astro-ph.GA},
    adsurl = {https://ui.adsabs.harvard.edu}
}

@ARTICLE{Vavilova2026b,
    author = {{Vavilova}, I.~B. and {Izviekova}, I.~O. and {Dmytrenko}, A.~M. and {Kompaniiets}, O.~V. and {Kukhar}, O.~N. and {Pastoven}, O.~S. and {Denyschenko}, S.~I. and {Dobrycheva}, D.~V. and {Fedorov}, P.~N. and {Khramtsov}, V.~P. and {Sergijenko}, O. and {Vasylenko}, A.~A.},
    title = "{Milky Way Near Twins (MWNeTs). II. Applications of a Hierarchical Framework to NGC 3521, NGC 2903, UGC 11487, and CGCG 047-026}",
    journal = {arXiv e-prints},
    year = 2026,
    month = oct,
    id = {arXiv:2610.23327},
    pages = {arXiv:2610.23327},
    doi = {10.48550/arXiv.2610.23327},
    archivePrefix = {arXiv},
    eprint = {2610.23327},
    primaryClass = {astro-ph.GA},
    adsurl = {https://ui.adsabs.harvard.edu}
}

@ARTICLE{Izviekova2026,
    author = {{Izviekova}, I.~O. and {Kompaniiets}, O.~V. and {Vavilova}, I.~B. and {Dobrycheva}, D.~V.},
    title = "{Revealing a multi-zone circumnuclear dust response in UGC 11487 through phase-resolved WISE W1--W2 diagnostics}",
    journal = {arXiv e-prints},
    year = 2026,
    month = jul,
    eid = {arXiv:2607.17815},
    pages = {arXiv:2607.17815},
    doi = {10.48550/arXiv.2607.17815},
    archivePrefix = {arXiv},
    eprint = {2607.17815},
    primaryClass = {astro-ph.GA},
    adsurl = {https://ui.adsabs.harvard.edu}
}

@ARTICLE{Kompaniiets2026,
    author = {{Kompaniiets}, O.~V. and {Vavilova}, I.~B. and {Vasylkivskyi}, Y.~V. and {Konovalenko}, O.~O. and {Pastoven}, O.~S. and {Izviekova}, I.~O. and {Junais} and {Dmytrenko}, A.~M. and {Dobrycheva}, D.~V. and {Fedorov}, P.~N. and {Khramtsov}, V.~P. and {Sergijenko}, O. and {Vasylenko}, A.~A.},
    title = "{NGC 3521 as the Milky Way near-twin: UV─radio-decameter spectral energy distribution}",
    journal = {\aap},
    year = 2026,
    month = may,
    volume = {710},
    eid = {A11},
    pages = {A11},
    doi = {10.1051/0004-6361/202558597},
    archivePrefix = {arXiv},
    eprint = {2512.14664},
    primaryClass = {astro-ph.GA},
    adsurl = {https://ui.adsabs.harvard.edu}
}

@ARTICLE{Yuan1969,
       author = {{Yuan}, C.},
        title = "{Application of the Density-Wave Theory to the Spiral Structure of the Milky way System. II. Migration of Stars}",
      journal = {\apj},
         year = 1969,
        month = dec,
       volume = {158},
        pages = {889},
          doi = {10.1086/150249},
       adsurl = {https://ui.adsabs.harvard.edu/abs/1969ApJ...158..889Y}
}

\begin{appendix}

\end{appendix}

\end{document}